\documentclass[twocolumn,floatfix,prb,aps,bibliography]{revtex4-1}
\usepackage{graphicx,color,url,nicefrac,color,multirow,amsmath,amssymb,amsfonts,graphicx,tabularx}
\usepackage[titletoc,title]{appendix}
\usepackage[unicode=true,colorlinks=true]{hyperref}
\hypersetup{linkcolor=blue,citecolor=blue,urlcolor=blue}
\makeatletter
\newcommand*{\rom}[1]{\expandafter\@slowromancap\romannumeral #1@}
\makeatother
\usepackage{xcolor}
\usepackage{mathtools}
\usepackage{graphicx,amsfonts,amsmath,mathrsfs,multirow,pbox,epstopdf,xfrac,color}
\begin{document}
\title{Cluster-cluster correlations beyond the Laughlin state}
\author{Bartosz Ku\'smierz $^{1,2}$ and Sreejith G J$^2$}
\address{$^1$ Department of Theoretical Physics, Wroclaw University of Science and Technology, Poland\\ $^2$ IISER Pune, Dr Homi Bhabha Road, Pune 411008, India}
\begin{abstract}
Number of zeros seen by a particle around small clusters of other particles is encoded in the root partition, and partly characterizes the correlations in fractional quantum Hall trial wavefunctions.
We explore a generalization wherein we consider the counting of zeros seen by a cluster of particles on another cluster. Numbers of such zeros between clusters in the Laughlin wavefunctions are fully determined by the root partition. However, such a counting is unclear for general Jain states where a polynomial expansion is difficult. 
Here we consider the simplest state beyond the Laughlin wavefunction, namely a state containing a single quasiparticle of the Laughlin state. 
We show numerically and analytically that in the trial wavefunction for the quasiparticle of the Laughlin state, counting of zeros seen by a cluster on another cluster depends on the relative dimensions of the two clusters.
We further ask if the patterns in the counting of zeros extend, in at least an approximate sense, to wavefunctions beyond the trial states. Using numerical computations in systems up to $N=9$, we present results for the statistical distribution of zeros around particle clusters at the center of an FQH droplet in the ground state of a Hamiltonian that is perturbed away from the $V_1$ interaction (short-range repulsion). 
Evolution of this distribution with the strength of the perturbation shows that the counting of zeros is altered by even a weak perturbation away from the parent Hamiltonian, though the perturbations do not change the phase of the system. 
\end{abstract}
\maketitle
\section{Introduction}
Trial wavefunctions in fractional quantum Hall effect (FQHE) are often characterized in terms of their vanishing properties. 
The Laughlin wavefunction at $\nu = 1/3$ vanishes as the third power of the distance between any pair of particles, which is stronger than what is required by the Pauli exclusion principle\cite{Laughlin83a}.
Equivalently, each particle coordinate in the wavefunction carries three zeros of every other particle coordinate. Holomorphicity of the wavefunction associates three vortices also to these three zeros. Pairs of particles in this state always have a relative angular momentum greater than or equal to three. The state is the densest zero energy state of a Hamiltonian which imposes an energy penalty on any pair of particles with relative angular momentum less than three.
These interrelated notions of the Laughlin state - namely electronic correlations as described by the vanishing properties, the parent Hamiltonian that enforces such electronic constraints, and the counting of zeros and vortices of particle coordinates - extend to a more general class of wavefunctions. The bosonic Pfaffian wavefunction vanishes as a second power of the mean distance between three nearby particles as they approach each other, but the wavefunction does not vanish when two of them come close. This property uniquely determines a parent Hamiltonian - namely a three body short-range repulsion - of which the Pfaffian is a unique incompressible ground state\cite{Moore91, Simon07a}.

These motivate us to consider a sequence $\lambda_n$, called the root partition, which enumerates the number of zeros that a particle sees in small clusters of $n$ particles. Set of all constraints encoded in this sequence uniquely determine the Laughlin state, Pfaffian state, and Read-Rezayi states\cite{Read99,Simon07a,Xiao008}. However, the same is not true in more general FQH trial states in the Jain sequence\cite{Jain89, Sreejith18, Regnault09}. It appears that one may need constraints more general than what can be specified using root partitions to uniquely determine the Jain states. 

A natural way to generalize the root partition is to consider the number of zeros, say $\gamma_{s,p}$ seen by a cluster of $p$ particles in a cluster of $s$ other particles. The root partition corresponds to the special case where $p$ is one. In Ref.~\onlinecite{Xiao008}, the authors use similar patterns of zeros approach to classify those symmetric polynomials of infinite variables in which the numbers $\gamma_{s,p}$ are well defined. These include the Laughlin, Pfaffian and Read-Rezayi states\cite{Xiao008,Read99}. Possible generalization to states in the Jain sequence  presents an appealing question. 

Simplest of the Jain sequence state beyond the Laughlin state is the state at filling $2/5$ of fermions. Composite fermion (CF) theory \cite{jain2007composite,Jain89} describes the wavefunction of this state as made of two filled Landau levels of CFs. 
Previous numerical studies on the zero-zero and zero-particle correlations in the Jain state at $2/5$ state suggest that the zeros effectively form a liquid with a short-range repulsion between them\cite{Graham03}.
An explicit expansion of this wavefunction to study zeros in clusters is, however, a challenging task computationally and analytically. We therefore consider the counting in a state that is a simpler step away from Laughlin state - namely the single quasiparticle (QP) of the Laughlin state. Such a state contains a fully filled first Landau level and a single composite fermion in an otherwise empty second Landau level of composite fermions. 

We analytically obtain the number of zeros $\gamma_{s,p}$ in such a wavefunction and show that the number depends on the sizes (radii) of the two clusters of particles. Numerical results in the case of few particle clusters agree with this analysis.
Having derived the properties in the trial wavefunction, we consider the question of how well these countings agree with the counting in states away from the trial wavefunctions. To this end, we consider the counting in the ground state and quasiparticle states of Hamiltonians perturbed away from the short-range repulsion $V_1$. We show that except in the simplest clusters  the counting of zeros is changed by even small perturbations as the zeros appear to be repelled away from the particles upon addition of perturbations. We present numerical results on the evolution of the distribution of zeros around simple clusters as the perturbation away from the parent Hamiltonian is tuned.

The paper is structured as follows.
In the next section, we present basic definitions used for the analysis of zeros attached to multi-particle clusters in FQH wavefunctions. 
In Sec.~\ref{sec3} we apply these ideas to the Laughlin ground state, single QP and single quasihole states.
Importantly, we show that the number of zeros attached to particle clusters as seen by a cluster with more than one particles is not uniquely determined by the root partition. We consider two kinds of quasiparticle trial wavefunctions, the CF QP and the Laughlin QP wavefunctions - which both have the same root partitions but the general counting of zeros are different.
Sec.~\ref{sec4} discusses the numerical results for zeros in the eigenstates of Hamiltonians which are perturbed away from the $V_1$ interaction Hamiltonian but fall in the same universality class as $V_1$ interaction. 

\section{Zeros of the FQHE polynomials}\label{sec:meth}
\subsection{Definitions}
\label{sec:defPolynomials}
FQHE wavefunctions in the LLL have a form
\begin{equation}\label{eq11}
\Psi(z_1,z_2\dots) \times \exp\left ({-\sum_i|z_i|^2/4} \right ),\nonumber
\end{equation} 
where $\Psi$ is an antisymmetric polynomial for the case of fermions and symmetric polynomial for  bosons. Throughout the manuscript we will use units where magnetic length is set to 1. The many particle wavefunction (Eq.~\ref{eq11}) is constructed from the single particle angular momentum eigenstates $z^m \exp(-|z|^2/4)$, where $m$ is the angular momentum of the particle. The information about correlations, is contained in the polynomial part; we therefore will not explicitly write the ubiquitous gaussian factor in the wavefunctions hereafter. Many features of the correlations are encoded in the distribution of zeros of this polynomial. In this section we introduce a few definitions regarding polynomials that are used in the rest of the paper.
   
A  partition is a sequence $\lambda=(\lambda_1, \lambda_2,\dots)$ of non-negative integers in non-decreasing order\footnote{A standard definition of partition\cite{macdonald1998symmetric,macdonald1998symmetric2,STANLEY198976} considers non-increasing order, However,here we use the opposite ordering}. Partitions can be ordered according to the  natural order, where we say that $\lambda$ dominates $\mu$ (denoted $\lambda \geq \mu$) if for every natural number $k$, the sum of the last $k$ parts of $\lambda$ is greater than or equal to the sum of the last $k$ parts of $\mu$.

The monomial symmetric functions (monomials) $m_{\lambda} (z_1, z_2, \dots, z_N)$ form a convenient basis for the symmetric polynomials and are defined as \cite{STANLEY198976,macdonald1998symmetric,macdonald1998symmetric2,Hora2007}
\begin{eqnarray}
m_\lambda &=& \mathcal{S}ym \left ( {z}_1^{\lambda_{1}} \times {z}_2^{\lambda_{2}} \times \dots \times {z}_N^{\lambda_{N}} \right ).\nonumber
\end{eqnarray}
Similarly, Slater determinants $\text{Sl}_{\mu}$ defined as \cite{macdonald1998symmetric,macdonald1998symmetric2}
\begin{equation}
\text{Sl}_{\mu} =\mathcal{A}nti\mathcal{S}ym \left ( z_{1}^{\mu_{1}} \times z_{2}^{\mu_{{2}}} \times \dots \times z_{N}^{\mu_{N}} \right ), \nonumber
\end{equation}
form a good basis for the antisymmetric polynomials. Monomial basis states are in one to one correspondence with the set of all partitions defined earlier. Slater determinants are labelled by a subset of all partitions which form a strictly increasing sequence of numbers (as opposed to non-decreasing)

Given a symmetric/antisymmetric polynomial $Q$ in $N$ variables, we may associate a root partition to it. A partition $\lambda$ is a root partition of a symmetric/antisymmetric polynomial $Q$ if its expansion has nonzero coefficients only for those monomials/Slater determinants labeled by partitions which are sub-dominant to $\lambda$ in natural order (and ${\rm Sl}_\lambda$ or $m_\lambda$ has a non-zero coefficient). 

Lastly, we define a translation symmetric polynomial as one that is symmetric under translations of all variables {\emph{i.e.},} $ Q(z_1,...z_N) =Q(z_1+a,...,z_N+a)$. Many quantum Hall ground state trial wavefunctions such as the CF states are translationally symmetric. Translation symmetric polynomials often can be associated with a root partition \cite{liptrap2010translation,haltalk}.

\begin{figure}[ht]
 \includegraphics[width=0.45\textwidth]{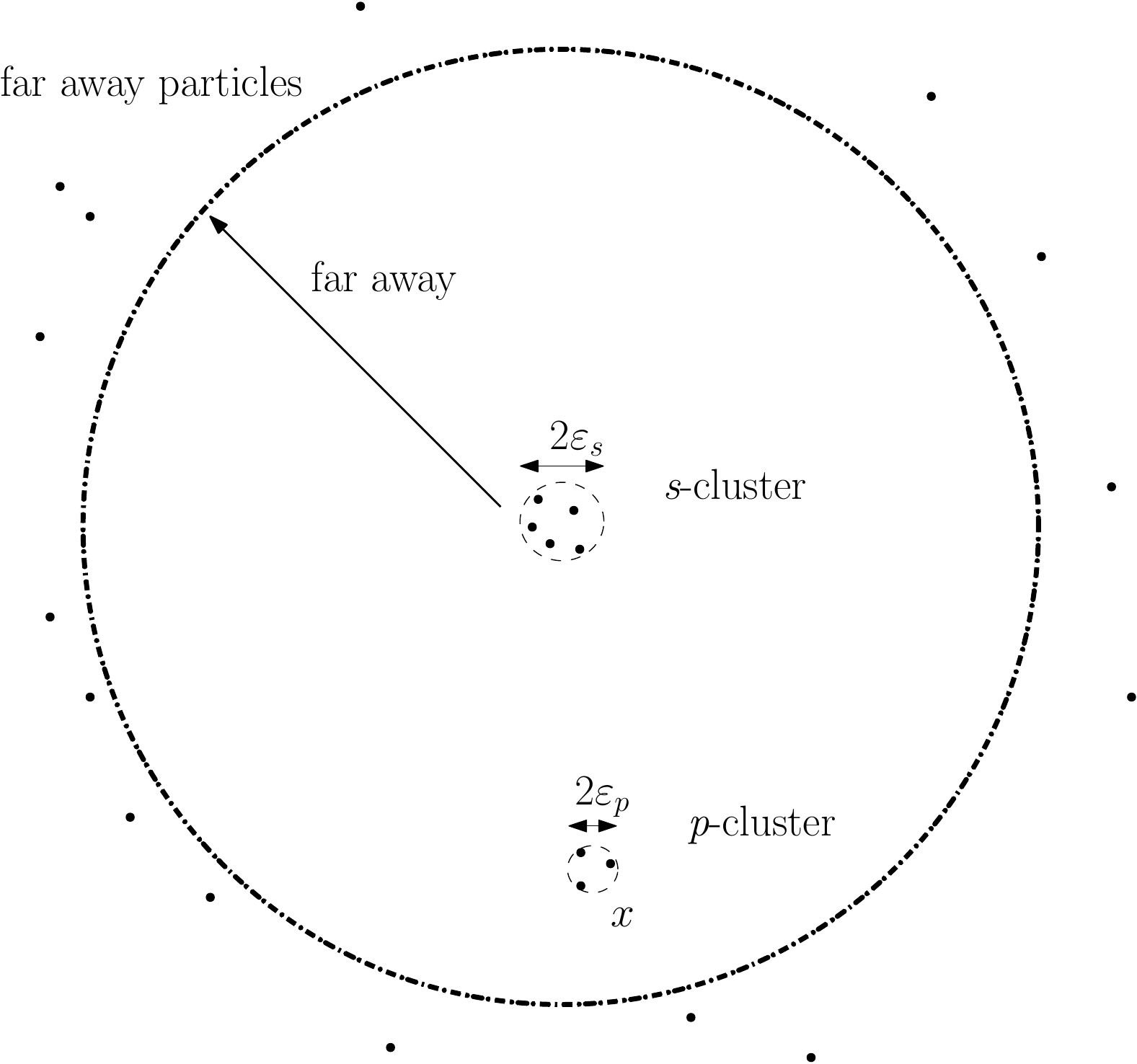} 
\caption{Configurations of particles showing the $s$-cluster, located here at the origin; $p$-cluster located around a position $x$ (which we treat as a variable); and far away particles, scattered at a large distance from the origin. }\label{fig:set}
\end{figure}

\subsection{Particle clusters}
We are interested in understanding the distributions of zeros around particles in the FQHE wavefunctions by numerically explicitly solving the polynomial equations. More specifically, we look at the distribution of zeros seen by a small cluster of particles, in the vicinity of another small cluster of particles. 
For an $N$ particle wavefunction on the disc, we divide the $N$ coordinates ($z_i$) into three groups: a cluster that ``probes'' the zeros ($p$-cluster), a static cluster ($s$-cluster), and the far away particles (consisting of $p$, $s$, and $f$ particles respectively, $N = p +s+f$).  

We consider configurations of particles similar to Fig.~\ref{fig:set} in which the particles in the $s$-cluster are located at $\varepsilon^{(s)}_i$, for $i=1,2,\dots s$, all within a small radius $\varepsilon_s$ around the origin {\emph{i.e.}}, maximum of $|\varepsilon_i^{(s)}|$ is of order $\varepsilon_s$
The far away particles, as the name suggests, are located far away from the static cluster. 
The $p$-cluster is centered around a location $x$ treated as a variable. The $p$ particles are located at positions $x+\varepsilon_i^{(p)}$, for $i=1,2,\dots p$ within a small radius of order $\varepsilon_p$  around $x$. We note that while we say that the $s$-cluster and the $p$-cluster are centered around the origin and $x$ respectively, we do not  require that they are the center of mass of the clusters.

We are interested in the counting of zeros of the wavefunction treated as polynomial in $x$ in the vicinity of the $s$-cluster. For translation symmetric polynomials, the counting of zeros around the $s$-cluster is independent of the location of the $s$-cluster (provided far away particles are still located away from the $s$-cluster). While the trial wavefunctions considered here are translation symmetric, this is not the case for eigenstates obtained from exact diagonalization of pseudopotentials. To be able to compare different cases, we will consider configurations where the $s$-cluster is centered at the origin.

\subsection{Reduced polynomials}\label{red_pol}
Consider a wavefunction of $N$ particles whose polynomial part is $\Psi(z_1,...z_N)$. 
When the positions of $s$-cluster and far away particles are fixed, and the positions of the particles in the $p$-cluster relative to its center $x$ are fixed, the wavefunction reduces to a polynomial  $P(x)$ in one variable, namely the center $x$ of $p$-cluster. 
For wavefunctions with a root partition $\lambda$, the degree of $P$ is given by $\sum_{i=1}^{p} \lambda_{N-i}$, which also equals the total number of zeros of $P$. We are interested in those zeros of this polynomial which fall on the $s$-cluster in the limit when the radii $\varepsilon_p$ and $\varepsilon_s$ of the $p$ and $s$ cluster are small.

\subsection{Zeros between clusters from the reduced polynomial} \label{secD}
Consider a configuration wherein the particles in the $s$-cluster are located at
\begin{equation}
\varepsilon_i^{(s)}=\varepsilon_su_i, {\text{ where }}i=1,2\dots s,\nonumber
\end{equation}
the particles in the $p$-cluster are located at
\begin{equation}
x+\varepsilon_i^{(p)}=x+\varepsilon_p v_i, {\text{ where }}i=1,2\dots p,\nonumber
\end{equation}
and the far away particles are located at $z_i$, $i=1,2\dots f$. $\varepsilon_p$ and $\varepsilon_s$ parameterize the radii of the two clusters and $u_i$ and $v_i$ have modulus of order $1$. The reduced polynomial in $P(x)$ which is given by 
\begin{equation}
\Psi(\varepsilon^{(s)}_1,\dots \varepsilon^{(s)}_s,x+\varepsilon^{(p)}_1,\dots,x+\varepsilon^{(p)}_p, z_{s+p+1}, \dots,z_N)\nonumber
\end{equation}
in general have a form 
\begin{equation}
\sum_{a,b,c} C_{abc} \varepsilon_s^a x^b \varepsilon_p^c \label{Eq:ps}
\end{equation}
where the coefficients $C_{abc}$ depend on the values of $z_i$, $u_i$ and $v_i$. The powers $a,b,c$ are non-negative integers.
We are interested in the roots of the polynomial in the limit where $\varepsilon_{p}$ and $\varepsilon_{s}$ are small.

\paragraph*{One particle in the $p$-cluster:} 
For the case where there is only one particle in the $p$-cluster, the polynomial has a simpler form 
\begin{equation}
\sum_{a,b} C_{ab} \varepsilon_s^a x^b
\end{equation}
In the spirit of $s$-cluster being small, we consider the limit where $\varepsilon_s$ approaches $0$. The polynomial is dominated by the term with the smallest $a$ such that $C_{ab}$ is non-zero. In other words, the $P$ approaches the following form
\begin{equation}
P(x)\to \varepsilon_s^{a_0} \sum_{b} C_{a_0b} x^b\label{Eqn:p1s}
\end{equation}
The number of zeros seen by $x$ at the origin is the smallest power $b$ in the above expression.

The exponent $a_0$ can be expressed in terms of the root partition, provided the wavefunction $\Psi$ has a well defined root partition. From the definition of the root partition given in Sec.~\ref{sec:defPolynomials},  it can be seen that $a_0$ is the sum of the first $s$ elements of the root partition  $a_0=\sum_{i=1}^s \lambda_i$. The number of zeros seen by $x$ at the origin is the next entry of the root partition, which is $\lambda_{s+1}$.

\paragraph*{More than one particle in the $p$-cluster:} 
If there are more than one particles in the $p$-cluster, then we need to consider a sequence of two limits where $\varepsilon_s$ and $\varepsilon_p$ go to 0. Consider the case where $\varepsilon_s\to0$ is performed first. Sending $\varepsilon_s$ to $0$ reduces $P$ to
\begin{equation} 
P\to \varepsilon_s^{a_0}\sum_{b,c} C_{a_0bc} x^b \varepsilon_p^c\label{Eq:psEpss0}
\end{equation}
where $a_0$ is smallest power $a$ of $\varepsilon_s$ in Eq.~\ref{Eq:ps} with a non-zero $C_{abc}$. Then we send $\varepsilon_p$ to 0 which brings the polynomial to
\begin{equation}
P\to \varepsilon_s^{a_0} \varepsilon_p^{c_0}\sum_{b} C_{a_0bc_0} x^b \label{Eq:psEpss0Epsp0}
\end{equation}
where $c_0$ is the smallest power of $\varepsilon_p$ in Eq.~\ref{Eq:psEpss0}. 

The number of zeros of $x$ at the origin is the smallest power of $x$ in Eq.~\ref{Eq:psEpss0Epsp0}. This is given by the smallest $b$ such that $C_{a_0 b c_0}$ is non-zero, wherein $a_0$ is the minimum number such that $C_{a_0 b c}$ is non-zero and $c_0$ is the minimum number such that $C_{a_0 b c_0}$ is non-zero.

If we reverse the order of limits, {\emph{i.e.}} send $\varepsilon_p\to0$ before sending $\varepsilon_s \to 0$  then the number of zeros seen by $x$ at the origin is different. The number of zeros of $x$ at the origin is given by the smallest $b$ such that $C_{a_0 b c_0}$ is non-zero, wherein $c_0$ is the minimum number such that $C_{a b c_0}$ is non-zero and $a_0$ is the minimum number such that $C_{a_0 b c_0}$ is non-zero. We now make four comments on the count of the zeros seen by $x$. 

\paragraph*{1. Cluster zeros cannot be obtained from the root partition:}
For a wavefunction with a well defined root partition, we saw that the number of zeros can be obtained from the root partition if the $p$-cluster had only a single particle. The argument can be phrased as follows: Merging of $n$ particles results in the wavefunction vanishing as $\gamma_n$ powers of the distance between them, where $\gamma_n=\sum_{i=1}^n\lambda_i$. Merging of $n+1$ particles results in the wavefunction vanish as a $\gamma_{n+1}$ powers of the distance between the particles. Thus the wavefunction should have vanished as $\gamma_{n+1}-\gamma_{n}=\lambda_{n+1}$ power of the distance between the $n$ particle cluster (which we interpret as the $s$-cluster) and the last particle (which we interpret as the $p$-cluster). This indicates that the $p=1$ particle cluster sees $\lambda_{n+1}$ zeros on the $s=n$ particle cluster.

It is tempting to generalize this to the case of more particles in the $p$ cluster. Merging of $p+s$ particles results in the wavefunction vanishing as $\gamma_{s+p}$ powers of the distances whereas merging of $p$ particles and $s$ particles separately is what resulted in $\gamma_p+\gamma_s$ of these powers. This suggests that the number of zeros seen by $p$-cluster on $s$-cluster is $\gamma_{p+s}-\gamma_s-\gamma_p$. 

Interestingly this turns out to be incorrect. The merger of $p$ particles does affect the power with which the wavefunction vanishes with distance upon the subsequent merger of the $s$ particles. The incorrect relation above is satisfied when the above long-range effects are ignored.   

\paragraph*{2. The order of limits $\varepsilon_s,\varepsilon_p\to 0$ matter:} Closely related to point {\it (1.)} made above is the fact that the two possible orders in which the limits are taken namely $\lim_{\varepsilon_s\to 0}\lim_{\varepsilon_p\to 0}$ and $\lim_{\varepsilon_p\to 0}\lim_{\varepsilon_s\to 0}$ yield, in general, different answers when $s\neq p$. We define $\gamma_{p,s}$ as the number of zeros seen by a $p$-particle cluster on the $s$-particle cluster given that $\varepsilon_p\to 0$ before $\varepsilon_s\to 0$. $\gamma_{s,p}$ represents the reverse ordering of the limits. The two orders of limits described above correspond to two paths along which $\varepsilon_p=\varepsilon_s=0$ can be reached. By choosing different path, for instance $\varepsilon_s=\varepsilon_p=\varepsilon$, we can get a range of possible answers as the number of zeros. We use a symbol $\gamma_{s}^{p} $ to denote the number of zeros seen by $p$-cluster on $s$ along the path $\varepsilon_p=\varepsilon_s\to 0$.

\paragraph*{3. Interchangeability of $s$ and $p$ in translation symmetric polynomials:} For a translation symmetric polynomial, the number of zeros seen by the $s$-particle cluster on a $p$-particle cluster is same as $\gamma_{p,s}$ provided the radius of the $p$-cluster is sent to 0 before sending the radii of $s$-cluster to $0$ {\emph{i.e.}}  patterns of zeros are invariant to $s\leftrightarrow p$ change.

\paragraph*{4. Non-translation symmetric polynomials:}
In general the coefficients $C_{abc}$ (Eq.~\ref{Eq:ps}) depend on the position of the far away particles, arrangement of the particles in the clusters, and position of the center of the $s$-cluster. For specific configurations of these, the specific term which decides the number of zeros, may have a coefficient $C_{a_0bc_0}$ which vanishes. In such rare configurations, the number of zeros may differ from the expected one. Small perturbations of the far away particles restore the coefficient and the counting.

It is easy to construct translation non-symmetric wavefunctions where this is not true. Simplest example is a quasihole wavefunction described in the next section. 
The important coefficient $C_{a_0bc_0}$ can turn out to be exactly zero if the positions of the $s$-cluster coincides with the quasihole. Changes to the positions of the far away particles do not affect this. In such cases the counting of the zeros may depend on the position of the central cluster.
Thus counting of the zeros of translation non-symmetric polynomials may in certain cases depend on the location of the $s$-cluster.


\section{Zeros between clusters in Laughlin state and excitations}\label{sec3}
In this section, we will analyze the zeros between clusters of particles in three FQHE trial states - Laughlin ground state, Laughlin quasihole and most importantly the Laughlin QP states. For the QP state, we consider two different trial wavefunctions, one proposed by Laughlin, obtained from the conjugation of quasihole creation operator\cite{PhysRevB.31.4026,Laughlin83} and the CF QP state \cite{jain2007composite}. 

\subsection{Laughlin $1/r$ state}\label{LAUGS} The Laughlin state \cite{Laughlin83a} at filling fraction $1/r$ (for positive integer $r$) is given by $\Psi_{\rm L}^{1/r} = \prod_{i<j} (z_i -z_j)^r$. Since the wavefunction depends only on differences between coordinates, the polynomial is translation symmetric; and the counting of zeros obtained for this wavefunction is independent of the location of the static cluster. Each of the $p$ particles of the $p$-cluster sees $r$ zeros on each of the $s$ particles of the $s$-cluster, therefore  
\begin{equation}
\gamma_{s,p} = \gamma_{p,s}= \gamma_{s}^{p}= rsp,
\end{equation} 
as was already discussed in Ref.~\onlinecite{Xiao008}. This matches the formula $\gamma_{s,p} = \gamma_{s+p} -\gamma_{s} -\gamma_{p}$.

\begin{figure}[ht]
\includegraphics[width=0.45\textwidth]{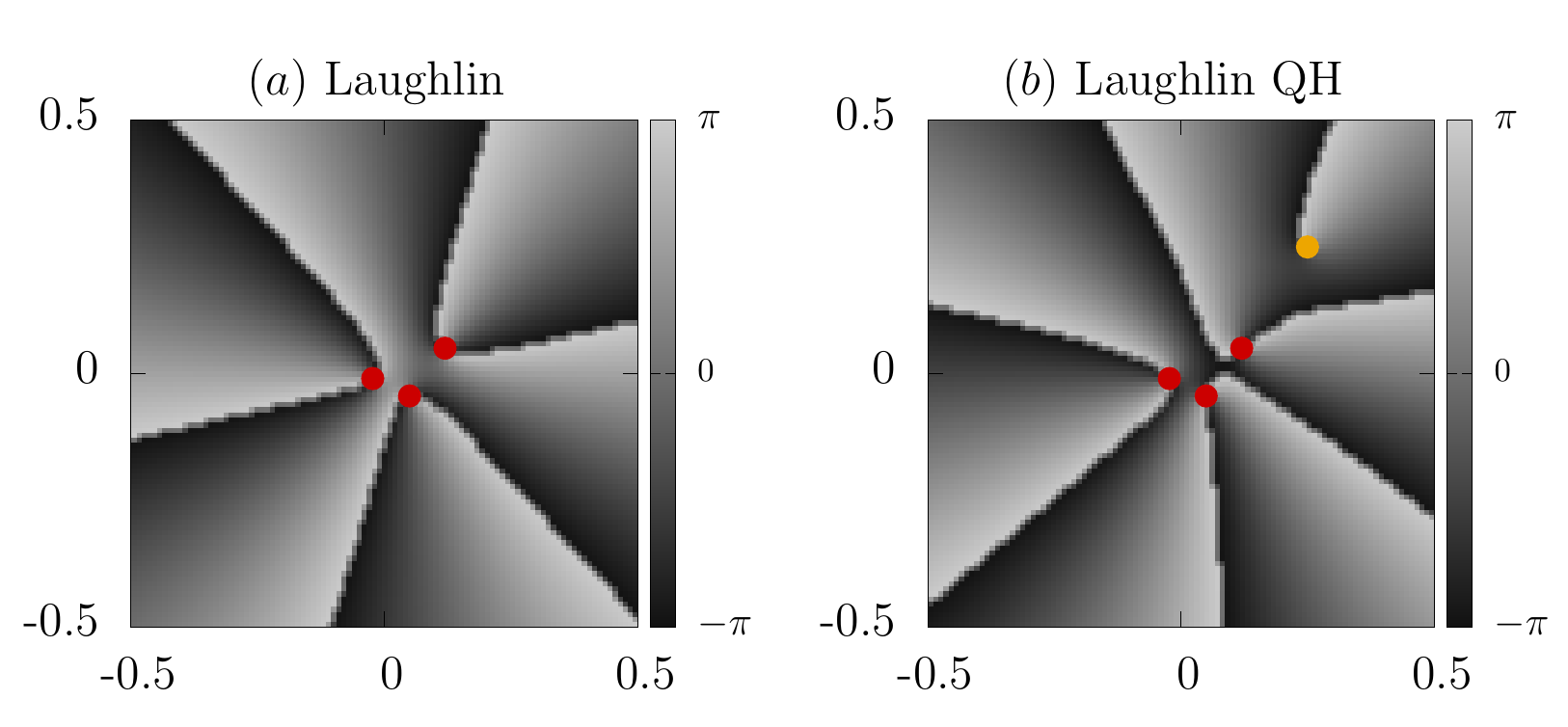} 
\caption{Illustration of zeros near a cluster of $s=3$ particles in trial wavefunctions at filing factor  $\nu=1/3$.  Arg-plot of $P(x)/J$ for $p=1$ particle in the $p$-cluster. Each zero acts as the source for a branch cut line across which the $\arg(P(x)/J)$ jumps from $-\pi$ to $\pi$.  The red dots are positions of particles and orange dot is a position of a quasihole. Panel $(a)$ shows the case of the Laughlin wavefunction, and panel $(b)$ shows the case of Laughlin quasihole wavefunction. Two non-Pauli zeros are located at each particle as indicated by two branch cuts originating each from particle. } \label{laughlin3}
\end{figure}

\subsection{Laughlin $1/r$ quasihole state}
A state with a single localized quasihole at $\eta$ over the Laughlin state at filling fraction $\nu = 1/r$ is described by the wavefunction
\begin{equation}\label{eq:qhh}
\Psi^{1/r}_{{\rm QH}} = \prod_{i} (z_i-\eta) \times \prod_{i<j} (z_i - z_j)^{r}.
\end{equation}
This state is not translation symmetric and the counting of zeros depends on the location of static cluster. The root partition is given by  $\lambda^{1/r}_{\rm QH} = (1,r+1,2r+1,3r+1,\dots)$, which is obvious for $\eta=0$ as multiplication by $\prod_i z_i$ maps every monomial/Slater determinant labeled by $\lambda = (\lambda_1,...\lambda_N)$ to the polynomial labeled by the partition $(\lambda_1+1,...\lambda_N+1)$. If $\eta \neq 0$ additional terms  appear, but their root partition would have only some of the elements increased by 1 and would be dominated by $\lambda^{1/r}_{\rm QH}$.

The reduced polynomial for this state and $p$ particles in the probing cluster can be calculated explicitly  
\begin{equation}\label{lauexp}
P(x) \propto \prod_{i=1}^{p} \left [ \left  (x+\varepsilon^{(p)}_i - \eta \right) \prod_{j=1}^{s} \left(x- \varepsilon^{(p)}_i + \varepsilon^{(s)}_j  \right )^r  \right ],
\end{equation}
where we show only that part of the full polynomial that is responsible for the zeros near the $s$-cluster and those near the quasihole. 

For $p$ particles in the probing cluster and for an $s$-cluster not located at the the position $\eta$ of the quasihole, the number of zeros equals $rsp$, same as in the ground state. Note that the $\gamma_{s,p=1}$ is not equal to the $(s+1)^{\rm th}$ element of the root partition. This is not surprising since the wavefunction is not translation symmetric.
If the $s$-cluster is located exactly at the position of quasihole, the quasihole contributes an additional $p$ zeros and the total number of attached zeros equals $rsp+p$. For $p=1$, this sequence matches the root partition.

As an illustration, Fig.~\ref{laughlin3} shows the color contour plot of $\arg(P(x)/J)$ for the Laughlin $\nu =1/3$ state. One Pauli zero is present on each particle in any fermionic wavefunction, so we have removed them to simplify the plots by dividing by the Jastrow factor $J=\prod_{i<k}(z_i-z_k)$. The zeros are the sources of the branch cuts of the arg function (the arg function jumps by $2\pi$ across these lines). Two non-Pauli zeros are located on each particle and one on the quasihole.

\subsection{Laughlin $1/r$ QP states}
Two different trial wavefunctions have been proposed to describe the localized QP wavefunction of the Laughlin state.

\paragraph*{1. Laughlin QP wavefunction:} The Laughlin QP has a structure similar to the wavefunction of the single localized quasihole, but with the $(z_i-\eta)$ factors in Eq.~\ref{eq:qhh} replaced with $(2\partial_{z_i}-\bar{\eta})$. Instead of adding an additional flux and thereby increasing the angular momentum of the particles around $\eta$, the Laughlin QP can be thought of as a localized anti-vortex which removes a unit of angular momentum around the $\eta$.

For a QP located at the position $\eta$ the Laughlin QP wave function can be written as \cite{Laughlin83,PhysRevLett.53.722,PhysRevB.31.4026} 
\begin{equation}\label{LQP}
\Psi_{\rm L}^{\rm QP} = \left (\prod_{k=1}^N  \left( 2\partial_{z_k} -\bar{\eta}  \right )\right ) \prod_{i<j} (z_i - z_j)^r.
\end{equation}

\paragraph*{2. Composite fermion wavefunction:} A better description of the QP wavefunction can be constructed using the composite fermion theory\cite{Jain89,jain2007composite,Jeon03}. A CF QP can be interpreted as a localized composite fermion in the second composite fermion Landau level.

The wavefunction of a single CF QP at the position $\eta$  is given by
\begin{equation}\label{CFunproj2}
\Psi_{\rm CF}^{\rm QP} =\mathcal{P}_{\rm LLL} \left (\Psi_{\rm CF}^{{\rm QP,unp}} \right),
\end{equation}
where $\mathcal{P}_{\rm LLL}$ is the LLL projection operator and the unprojected CF QP wavefunction is given by
\begin{equation}\label{CFunproj}
\Psi_{\rm CF}^{\rm QP,unp} =\left (\sum_i \frac{(\bar{z_i}-\bar{\eta})e^{\left ( \frac{\bar{\eta z_i}}{2\ell^2}-\frac{|\eta|^2}{4\ell^2}\right)}}{\prod_{{j\neq i} } (z_i -z_j)}\right ) \prod_{i<j}(z_i - z_j)^r.
\end{equation}

The orthogonal projection $\mathcal{P}_{\rm LLL}$  needed in the Eq.~\ref{CFunproj2} can implemented by replacing of the complex conjugated variables in unprojected state with partial derivatives $\bar{z}_i \to 2\partial_{z_i}$ \cite{doi:10.1142/S0217979297001301} to get
\begin{equation}\label{CFproj}
\begin{split}
    \Psi_{\rm CF}^{\rm QP}&= \prod_{i<j}(z_i-z_j)^{r-1}\times \\
    &\times \mathcal{A}nti\mathcal{S}ym \Bigg [ \left(\frac{1}{z_{1}-z_{2}}\right)\prod_{\substack{i<j\\2 \leq i,j }}^{N}\left(z_{i}-z_{j}\right)\Bigg].
\end{split}
\end{equation}

\subsubsection{Translation invariance of QP sates}
Now we consider translation invariance of the wavefunctions for the QP located at  $\eta=0$. Surprisingly, even though the QP is created at the origin both Laughlin QP and CF QP wavefunctions are translation symmetric. Translation symmetry of projected CF state is easy to see as the wave function (Eq.~\ref{CFproj}) depends only on the relative differences of particles' position. 

The Laughin QP state is translation symmetric because the derivatives in Eq.~\ref{LQP}, commute with translations of all of the variables by a common constant, and the Jastrow factor is a function of differences of coordinates.

Even the the unprojected CF state (Eq.~\ref{CFunproj2}) is translation symmetric, because the only term that does not depend on the relative positions is $\bar{z}_i$ in the numerator  of (Eq.~\ref{CFunproj2}) (for $\eta=0$ exponent factor equals 1). When all of the particles are shifted by a constant $a$, the the entire wave function changes by $ \left (\sum_i \frac{\bar{a}}{\prod_{{j\neq i} } (z_i -z_j)}\right ) \prod_{i<j}(z_i - z_j)^r$ which equals zero. 



In the following subsection we will analyze the the zeros between clusters of particles in these trial wavefunctions.

\subsubsection{Root partitions of QP states}
\paragraph{Laughlin QP at the origin:} The root partition of the Laughlin QP state (Eq.~\ref{LQP}) with a QP located at 0, can be obtained, by applying the product of partial derivatives to each of the monomials/Slater determinant appearing in the expansion of the $r^{\rm{th} }$ power of the Jastrow factor (see Eq.~\ref{LQP}). The operator $\prod_i \partial_{z_i}$ annihilates every monomial/Slater determinant in which one or more particle occupies the zero angular momentum orbital. The remaining monomials/Slater determinants, get one angular momentum removed from each of the occupied orbital. The highest partition in the expansion of $J^r$ (in natural order) with no non-zero elements is given by $(1,r-1,2r,3r,...,r(N-1))$. When one angular momentum is removed from each orbital we get the root partition of Laughlin QP state 
\begin{equation}\label{parti1}
\lambda_{\rm QP}^{1/r} =(0,r-2,2r-1,3r-1,...,r(N-1)-1). 
\end{equation}

\paragraph{CF QP at the origin:}
We can also calculate the root partitions for the CF $\nu=1/r$ state with $n$ composite fermion QPs located in successive momentum orbitals, using methods similar to that in the Ref.~\onlinecite{Sreejith18}. Here, we summarize the argument for just $1$ QP located in the lowest momentum orbital. The wave-function can be written as
\begin{multline}\label{projnew}
\Psi_{\rm CF}^{\rm QP} = \prod_{i<j}^N (z_i-z_j)^{r-2} \times \sum_{P\in S_N}\sum_{Q\in S_N} (-1)^{PQ}\\ \left[z_{P(1)}^0z_{P(2)}^1\dots z_{P(N-1)}^{N-2} \partial_{z_{P(N)}}\right]\times\\ 
\left[z_{Q(1)}^0z_{Q(2)}^1\dots z_{Q(N-1)}^{N-2} z^{N-1}_{Q(N)}\right]
\end{multline}
We will first focus on the terms arising from the double summation over permutations $S_N$. The sum is a symmetric function and therefore has an expansion in monomials.

Terms in the first bracket may be associated with angular momentum sequence $(0,1,\dots N-2,-1)$ and the ones in the second bracket can be associated with the angular momentum sequence $(0,1,2\dots N-1)$. Note that the derivative effectively acts as if it has a negative momentum because it reduces the momentum by 1 of the function on which it acts. However, we have to keep in mind that action of $\partial_{z_i}$ on a single particle state with 0 momentum ({\emph{i.e.}}~$z_i^0$) produces 0 and annihilates the term. Upon summing over permutations, the two brackets act in every possible permutations.

In summary, the partitions labeling the monomials/Slater determinants appearing in the expansion of the summation can be obtained by considering all possible ways in which a permutation of $(-1,0,1,2,3,\dots (N-2))$ can be added to $(0,1,2,\dots N-1)$ keeping in mind that any case with $-1$ in the sum is to be ignored as these represent action of the derivative on the $0$ momentum state. From this, it is easy to see that the dominant partition among these is obtained upon adding $(0,-1,1,2,\dots N-2)$ to $(0,1,2\dots N-1)$ which gives $(0,0,3,5,7,9\dots)$.

Now multiplication of this polynomial by $r-2$ powers (Eq.~\ref{projnew}) of the Jastrow factors merely adds $r-2$ zeros to every particle. This increments the number of zeros seen by a particle on an $s$-particle cluster by $s(r-2)$. Incorporating this makes the root partition the same as partition in Eq.~\ref{parti1}. Thus CF QP and Laughlin QP states have the same root partition.

\begin{figure}[ht]
\includegraphics[width=0.45\textwidth]{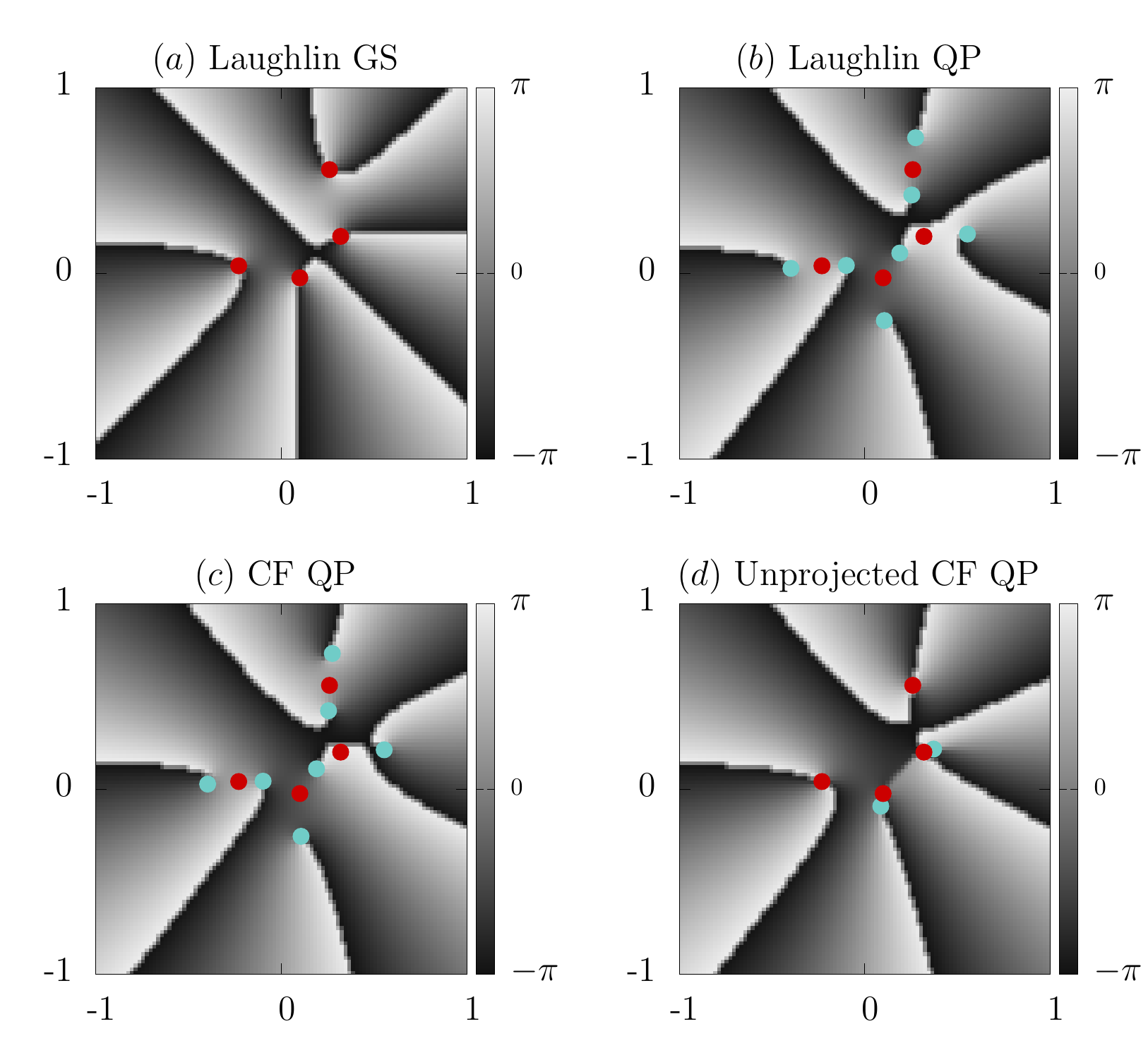} 
\caption{Illustration of zeros near the cluster of $s=4$ particles in trial wavefunctions at filing factor  $\nu=1/3$.  Arg-plot of $P(x)/J$ for $p=1$ particle in the $p$-cluster. The red dots are positions of particles and blue points are the positions of zeros, each zero acts as the source for a branch cut line. $(a)$ is the Laughlin ground state, $(b),(c)$ and $(d)$ are QP states: Laughlin QP,  CF QP and, unprojected CF trial wavefunctions respectively. In the unprojected CF trial wavefunction, which is not holomorphic $(d)$ some of the zeros may have opposite phase winding.   }\label{fig:QP1}
\end{figure}

\subsubsection{Illustration of zeros in QP states}
Fig.~\ref{fig:QP1} shows the arg-plot of $P(x)/J$ showing the zeros seen by a single particle $x$ in an $s=4$ particle cluster for the case of different wavefunctions. The locations of the four particles are kept the same in each case for comparison. Note that the Pauli zero has been removed from each particle by dividing the reduced polynomial by the Jastrow factor $J$.

The Laughlin state at $\nu =1/3$ (Fig.~\ref{fig:QP1} {$(a)$}) has two non-Pauli zeros located on each of the particles, resulting in 8 non-Pauli zeros in total in the $s$-cluster. 

Zeros are located away from the particles in the Laughlin and the CF QPs (Fig.~\ref{fig:QP1}{ $(b),(c)$}). The root partitions of the Laughlin and CF QPs are the same and there are seven zeros in the four particle cluster as expected from the root partition shown in Eq.~\ref{parti1}. Note that in the limit where the size $\varepsilon_s$ of the $s$-cluster approaches $0$, the zeros of $x$ as well as the four particles will converge to the same point. 

Unlike the zeros of projected wavefunctions, the zeros of unprojected CF state cannot be understood using similar techinques, as it contains conjugated variables and locating positions of all of the zeros is a tedious task. The unprojected CF wave function has always at least $r$ zeros on each particle, but there can be several additional zeros in the cluster depending on the precise configuration of the particles as shown in Fig.~\ref{fig:QP1} $(d)$. Since the additional zeros are related to the $\bar{z}$ variables they can have the opposite phase winding. Note also that the number of zeros in the states with $\bar{z}$ can not be obtained from counting the branch cut lines leaving some area (which is the case of polynomials in $z$ only). See appendix \ref{unprr} for more discussion of the unbounded number of zeros in unprojected CF wavefunction for symmetrical configurations of particles.

\subsubsection{Zeros between clusters in the CF QP state} \label{CFzeross}
The number of zeros $\gamma_{s,p}$ between clusters with $p \geq2$, $s \geq 2$ was fully determined by the root partition in the case of the Laughlin state. However, the same is not true for the CF QP wavefunction. 
A naive counting of the zeros from the root partition $\lambda_{\rm QP}^{1/r} = (0,r-2,2r-1,3r-1,4r-1,...)$ gives 
$\gamma_{s+p}-\gamma_{s}-\gamma_{p}$ which evaluates to $rsp$. This however, does not give the correct counting.

To find the correct number of zeros we analyze the Eq.~\eqref{CFproj}. The wavefunction contains a Jastrow factor part, and a part that involves an antisymmetrization.
We can find the counting of zeros by following the procedure given in section \ref{secD}. 

Consider the two clusters of with $s \geq 2$ and $p \geq 2$ particles. We will first calculate the number of zeros when the $\varepsilon_s$ is sent to zero before $\varepsilon_p$ is sent to zero.
The wavefunciton decays as $\varepsilon_s^{a_0}$ as the size $\varepsilon_s$ of the $s$-cluster approaches zero, as discussed in Sec.~\ref{secD}.
$a_0$ can be determined from the above wavefunction by inspection. 
The contribution to $a_0$ from the Jastrow factor is easily found to be $(r-1)s(s-1)/2$.
The contributions to $a_0$ from the remaining part of Eq.~\ref{Eq:psEpss0} arise from the terms in the expansion of the antisymmetrization which decays the slowest. These correspond to the terms in which both $z_1$ and $z_2$ shown in Eq.~\ref{CFproj} are in the $s$-cluster. Contribution to $a_0$ from such terms can be found to be $(s-1)(s-2)/2 -1$. Combining these, we get that $a_0 = \frac{1}{2}s(rs-r-2)$. 

Having fixed the value of $a_0$ we send the size $\varepsilon_p$ of the $p$-cluster located at $x$ to $0$.
Since the wavefunction is dominated by terms in which with $z_1$ and $z_2$ both belonging to $s$-cluster, the coordinates in $p$-cluster should be from $z_k$ for $k>2$ in the form of the wavefunction written in Eq.~\ref{CFproj}. 
These vanish with $\varepsilon_p$ as is  $p(p-1)/2^{\rm{th}}$ power. Combining the contribution from the Jastrow factor we find that the wavefunction vanishes as $rp(p-1)/2^{\rm{th}}$ power of $\varepsilon_p$ if $\varepsilon_p$ is sent to zero after $\varepsilon_s$ is sent to zero. 
Comparing with the discussion in Sec.~\ref{Eq:psEpss0} the wavefunction vanishes as 
\begin{equation}
    \varepsilon_s^{a_0} \varepsilon_p^{c_0} \sum_b C_{a_0 b c_0} x^b \label{eq:ref1} 
\end{equation}
where $a_0=\frac{1}{2}s(rs-r-2)$ and $c_0=rp(p-1)/2$
From the discussion in Sec.~\ref{secD} we know that if all of the $p+s$ particles are sent to a common point, the wavefunction scales as the following power of the relative distance.
\begin{equation}
\sum_{i=1}^{p+s} \lambda_{i}
\end{equation}
which for the root partition of the QP state can be found to be $rps$. From Eq.~\ref{parti1} we can estimate the same by setting $\varepsilon_s=\varepsilon_p=x\propto\varepsilon$. We find that the wavefunction scales $\varepsilon$ with a power  $a_0+c_0+\min(b)$ where $\min(b)$ is the smallest $b$ such that $C_{a_0 b c_0}$ is non-zero. Comparing the two estimates we find that $\min(b)$ is $rps-p$. The number $\min(b)$ is also the number of zeros seen by the $p$-cluster on the $s$-cluster. Thus we can infer that $\gamma_{s,p}=rps-p$.

A similar calculation in the scenario where $\varepsilon_p$ is sent to $0$ before $\varepsilon_s$ gives $\gamma_{p,s}=rps-s$. Similar considerations also tell us that if both $\varepsilon_p=\varepsilon_s=\varepsilon$ then the wavefunction vanishes as $\gamma^p_s=rsp-\min(p,s)$ where $\min(p,s)$ is the minimum of $p,s$.

\begin{figure}[ht]
\includegraphics[width=0.45\textwidth]{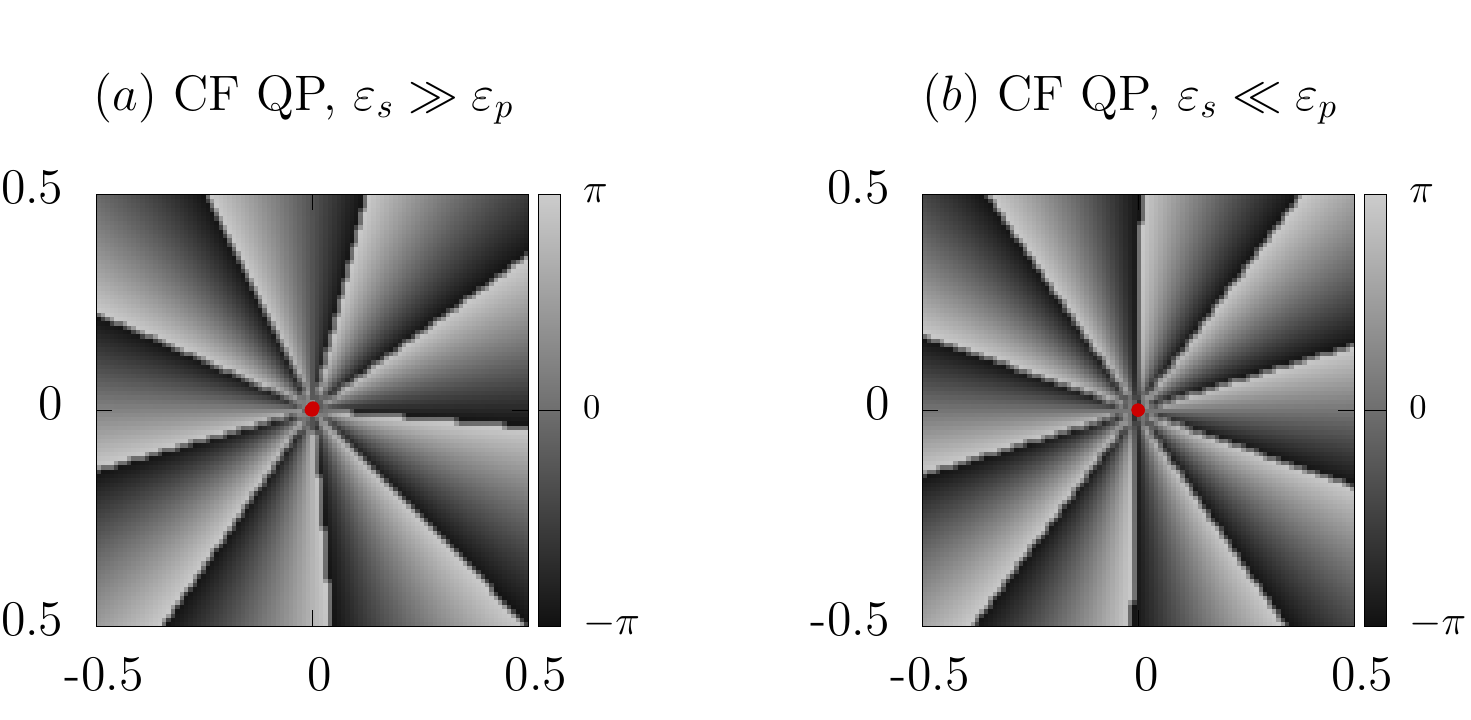} 
\caption{Illustration of the number of zeros seen by a $p=2$ particle cluster near the cluster of $s=3$ particles in the CF QP state. Figure shows the arg-plot of $P(x)/J$ for CF QP state at $\nu = 1/3$. Each zero acts as a source of branch cut line across which the $\arg(P(x)/J)$ jumps from $-\pi$ to $\pi$. In the panel $(a)$ $\varepsilon_p \ll \varepsilon_s$ results in 9 zeros, whereas in the panel $(b)$ $\varepsilon_s \ll \varepsilon_p$ results in 10 zeros. Observed number of zeros (which can be found by counting the branch cut lines) match calculations from \ref{CFzeross}. 
\label{2+3fig}}
\end{figure}

Interestingly, the different counting of zeros in CF QP states manifest in the reduced polynomials of finite cluster diameters, if one of radii is significantly smaller than the other. This is illustrated in the Fig.~\ref{2+3fig} where we plot the arg plot of CF QP  wavefunction at $\nu = 1/3$ for $s=3$ and $p=2$ (only non-Pauli zeros). The manipulation of the cluster diameters leads to the different number of zeros near the origin which equal $\gamma_{3,2}-6 = 9$ in part $(a)$ or $\gamma_{2,3}-6 = 10$ in part $(b)$ (note that $6$ zeros Pauli were subtracted).

\begin{figure}[ht]
\includegraphics[width=0.45\textwidth]{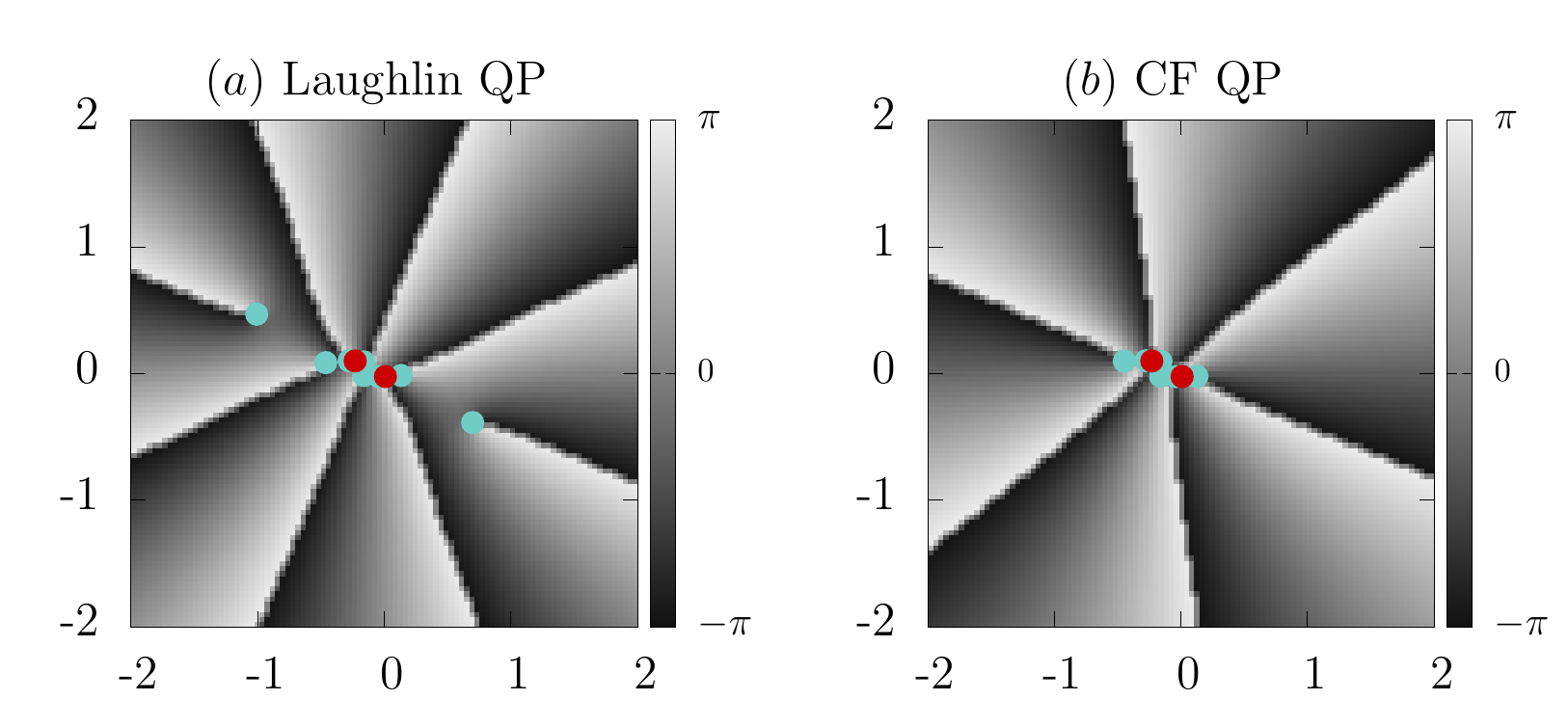} 
\caption{Illustration of different number of zeros seen by a cluster of $p=2$ particles near the cluster of $s=2$ particles. Figure shows the arg-plot of $P(x)/J$  for two trial QP wavefunctions at filling factor $\nu =1/3$.  Each zero acts as a source of a branch cut line across which the $\arg(P(x)/J)$ jumps from $-\pi$ to $\pi$. Panel $(a)$ is the Laughlin QP wavefunction and panel $(b)$ is the CF QP trial wavefunctions. The red dots are the positions of particles and blue dots are zeros. Despite having the same root partition, two trial wavefunctions have different number of zeros. }\label{fig:laqppp}
\end{figure}

An analysis of Laughlin QP state is much more complicated and we were not able to provide analytical calculations. However, we confirmed numerically that the number of zeros can be obtained from the partition and  $\gamma_{p,s}=\gamma_{s,p} =\gamma_{p+s}-\gamma_s-\gamma_p$ for cluster sizes up to $s,p \leq 4$. Here, we illustrate different number of zeros in the Fig.~\ref{fig:laqppp} where $s=p=2$. The total number of non-Pauli zeros at $\nu = 1/3$ for the Laughlin QP state equals $8$ and $6$ for CF QP. Further evidence is presented in Sec.~\ref{ref4furt}.

\section{Zeros between clusters in exact diagonalization states} \label{sec4}

\begin{figure}[ht]
\;\;\;\;\;\;\;\;\;\; \includegraphics[width=0.35\textwidth]{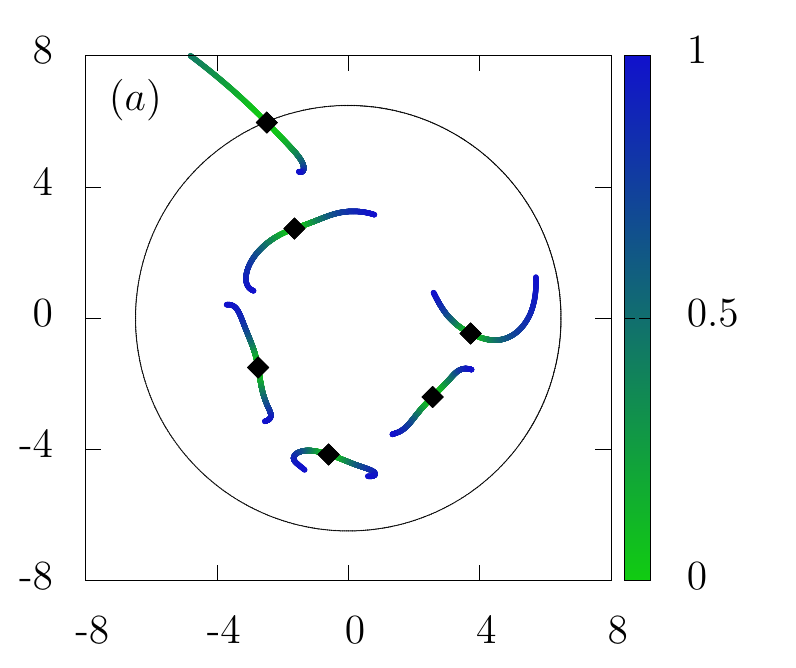} \\\vspace{0.25cm}
\includegraphics[width=0.40\textwidth]{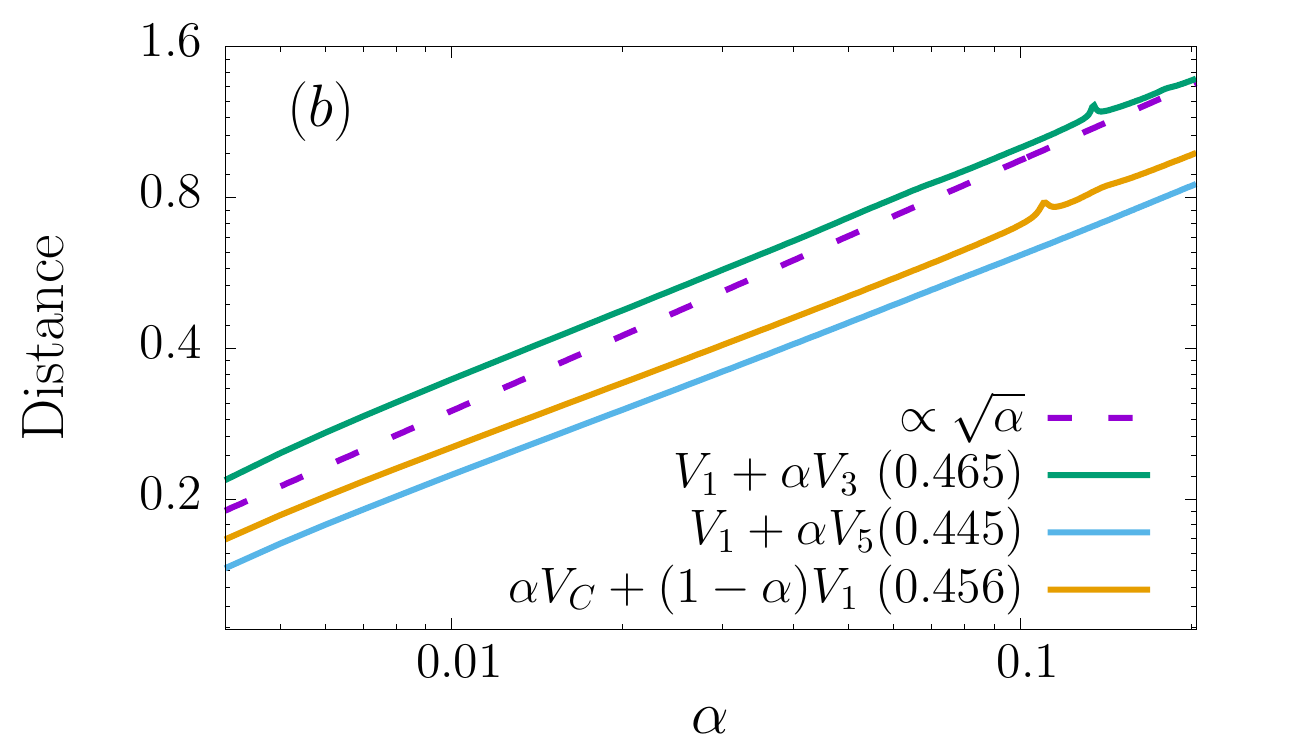} 
\caption{Decoupling of zeros from particles at $\nu = 1/3$ for different model interactions. Panel $(a)$ shows the zeros in one particular configuration of particles. Black diamonds mark the particles. The two lines attached to the particle locations show how the zeros drift away from the particle as $\alpha$ which parametrizes the strength of the perturbation away from $V_1$ is increased from $0$ to $1$. The color of the line represents the strength $\alpha$.
The large black circle shows the size of the droplet with radius $\sqrt{2 N/ \nu}$, $N=7$. 
Panel $(b)$ shows an average mean distance between zeros and particle (log-log scale) for $\alpha V_C +  (1-\alpha) V_1$, $V_1 +  \alpha V_3$ and $V_1 + \alpha  V_5$ interactions. In the legened, next to the interaction type are the exponents of fit to $distance = \alpha^k$.   }\label{zerpaytth}
\end{figure}

\subsection{Model pseudopotentials}

Any two-body Hamiltonian in LLL on a disk has the following form
\begin{equation*}
    V \equiv \sum_{i} V(i) P_i,
\end{equation*}
where $P_i$ is an operator that projects pairs of particles to a state of relative angular momentum $i$. $V(i)$ are the Haldane pseudopotentials, which specify the energy cost of a pair of particles being in the angular momentum\cite{Haldane83,Quinn00,PhysRevB.97.245125,PhysRevB.34.2670}. These pseudopotentials uniquely parametrize rotation symmetric symmetric two particle interactions in the LLL. 

The short-range repulsion interaction, represented by $V_1$, has a non-zero pseudopotential only at angular momentum $1$, thus it is defined by $V_1(1) = 1$ and $V_1(m\geq 3) =0$. 
Laughlin state at filling $\nu = 1/3$ is the densest zero energy ground state of this interaction\cite{Haldane83}. 

As discussed in Sec.~\ref{LAUGS}, every particle in the Laughlin state sees $3$ zeros located on all other particles. 
Upon perturbing away from the $V_1$ pseudopotential and towards the Coulomb interaction, the ground state remains in the same phase, however, the three zeros do not stay on the particle anymore. 
%

It is natural to ask if the number of zeros on the particle remains the same in an approximate sense as we perturb away from the $V_1$ point but staying within the same phase.
More generally we ask if the general zero counting between clusters holds true for the ground state and the QP state.

The $V_1$ interaction can be perturbed along various directions, however, we mainly consider perturbations towards the Coulomb $V_C$ interaction described by the pseudopotentials
$V_C(m) =\frac{2}{\sqrt{\pi}} \frac{\Gamma(m+1/2)}{\Gamma(m+1)}$. The transformation of the interaction from $V_1$ to $V_C$ can be parametrized linearly as
\begin{equation}
V^{\alpha} = \alpha V_C +  (1-\alpha)V_1. 
\end{equation}
where $\alpha=0$ corresponds to the $V_1$ interaction and $\alpha=1$ corresponds to the Coulomb interaction.

\subsection{Numerical methods}\label{nummet}
In order to numerically find the number of zeros on an $s$-cluster we need to evaluate the wavefunction for configurations of particles with particles close together. The wavefunctions vanish to $0$ rapidly if the sizes of the $s$ and $p$-clusters are sent to $0$. For reliable numerical calculations, we need to keep the particles in the two clusters at some finite radii $\varepsilon_s$ and $\varepsilon_p$. As can be seen in Fig.~\ref{fig:QP1}, if the radii of the cluster are finite, zeros which should ideally be situated at the centers of the $s$-cluster, instead stay in a finite neighbourhood arround the the $s$-cluster. Moreover, when we consider states other than the trial wavefunctions, for instance, the ground states of $V^\alpha$ interactions, we do not expect that the zeros strictly approach the center of the $s$-cluster even in the $\varepsilon_s,\varepsilon_p\to 0$ limits. In such cases the precise location of the zeros can be dependent on the coordinates of the far away particles as well as the positions of the particles in the $s,p$-clusters relative to their centers. 

As a result, analysis of a specific configuration of particle locations may not be sufficient to understand the distribution of zeros. To remedy this, we consider the statistical distribution of zeros averaged over an ensemble of configurations of particles. The most natural ensemble to consider is the distribution defined by the wavefunction itself. 

To construct a configuration with $s$ particles located in the vicinity of origin, we use Monte Carlo to sample the distribution
\begin{multline}
p(\theta_1,\theta_2,\dots,\theta_s,z_{s+1},z_{s+2}\dots z_N) =\\
\left| \Psi(\varepsilon_s e^{\imath \theta_1},\varepsilon_s e^{\imath \theta_2},\dots, \varepsilon_s e^{\imath \theta_s},z_{s+1},z_{s+2}\dots z_N)  e^{{-\sum_i|z_i|^2/4}} \right|^2
\end{multline}
where $\varepsilon_s$ is set to a number much smaller than the typical inter particle distances. For our calculations we chose $\varepsilon_s\sim 0.1$ in magnetic length units.

Every time a sample configuration $(\varepsilon_s e^{\imath \theta_1},\varepsilon_s e^{\imath \theta_2},\dots, \varepsilon_s e^{\imath \theta_s},z_{s+1},z_{s+2}\dots z_N)$ is obtained from Monte Carlo, $p$ of the coordinates from $z_{s+1},z_{s+2},..., z_{N}$ are replaced with $x+\varepsilon_p v_i$ where $x$ is a taken to be a variable and $v_i$ is a random complex number of modulus $1$. From this configuration, the reduced polynomials $P(x)$ are computed as described in Sec.~\ref{red_pol}.

It is computationally expensive to use Monte Carlo methods to sample points from the wavefunctions from exact diagonalizations. This is because these involve a large number of Slater determinant evaluations. In order to simplify the calculations, we sampled the configurations using the closest available trial wavefunctions that are easy to compute. For instance, while constructing the distributions of zeros in the ground states of the $V^\alpha$ interactions, we used the Laughlin state to sample the points. While studying the QP states of $V^\alpha$, we used the CF QP wavefunction to sample the configurations.

The zeros are located using the companion matrix method, which identifies the roots and the eigenvalues of the companion matrix. This method is sufficiently fast and accurate for considered system sizes\cite{10.2307/2153450}.

\subsection{Zeros decoupling from particles}

\begin{figure}[ht]
 \includegraphics[width=0.45\textwidth]{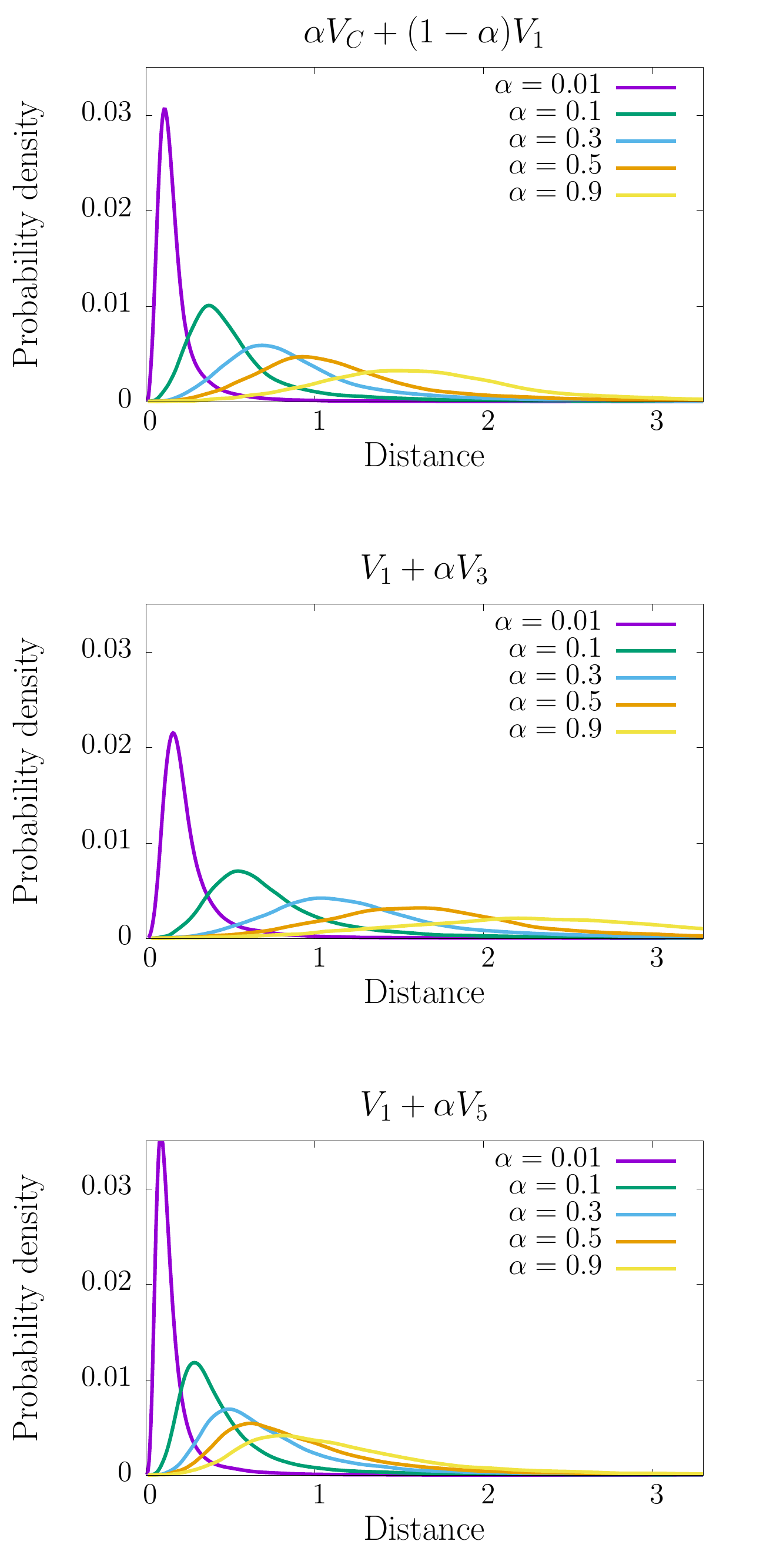} 
\caption{The probability density of finding a zero at a given distance from the particle from which the zero decoupled. Different lines show the distribution for different strengths of the perturbation $\alpha$ away from $V_1$. Different panels show the distribution for different perturbations $\alpha V_C + (1-\alpha) V_1$, $V_1 + \alpha V_3$ and $V_1 + \alpha V_5$. 
}\label{track}
\end{figure}

The ground state of the interaction $V^{\alpha}$ at $\alpha=0$, is the Laughlin state in which every particle sees three zeros on any other particle. As $\alpha$ is increased from $0$, the ground state of $V^\alpha$ still has three zeros in the vicinity of every particle and one of them is always located on the particle. For a given configuration of the $N-1$ particles $(z_1,z_2,z_3\dots z_{N-1})$, the zeros of $P(x)=\Psi_\alpha(z_1,z_2,\dots z_{N-1},z_N=x)$ where $\Psi_\alpha$ is the ground state of the $V^\alpha$ interaction, evolves with $\alpha$ continuously. We show this in Fig.~\ref{zerpaytth} $(a)$. The color scale indicates $\alpha$. We find that at small $\alpha$, the zeros drift symmetrically along diametrically opposite directions.

In Fig.~\ref{zerpaytth} $(b)$ we show the mean distances between each particle and the zeros drifting away as a function of $\alpha$. We consider three different perturbations away from the $V_1$ interactions, namely the  Coulomb $V_C$, $V_1+V_3$ interaction, and $V_1+V_5$ interactions, where $V_3(3)=1$ and $V_3(m\neq 3)=0$ and $V_5$ is defined similarly. Mean distance is obtained by averaging over different realizations of the configurations. Irrespective of the nature of the perturbation ($V_C,V_1+V_3$ or $V_1+V_5$), we find that the distance scales approximately as $\alpha^{1/2}$.

In Fig.~\ref{track}, we show the probability distribution of the distances of the zeros from the particles for three different interactions. In each case, as $\alpha$ is increased the zeros on an average drift away from the particle. The $V_3$ perturbation causes a greater repulsion of the zeros as compared to the $V_5$ interaction.

\begin{figure*}
\includegraphics[width=0.95\textwidth]{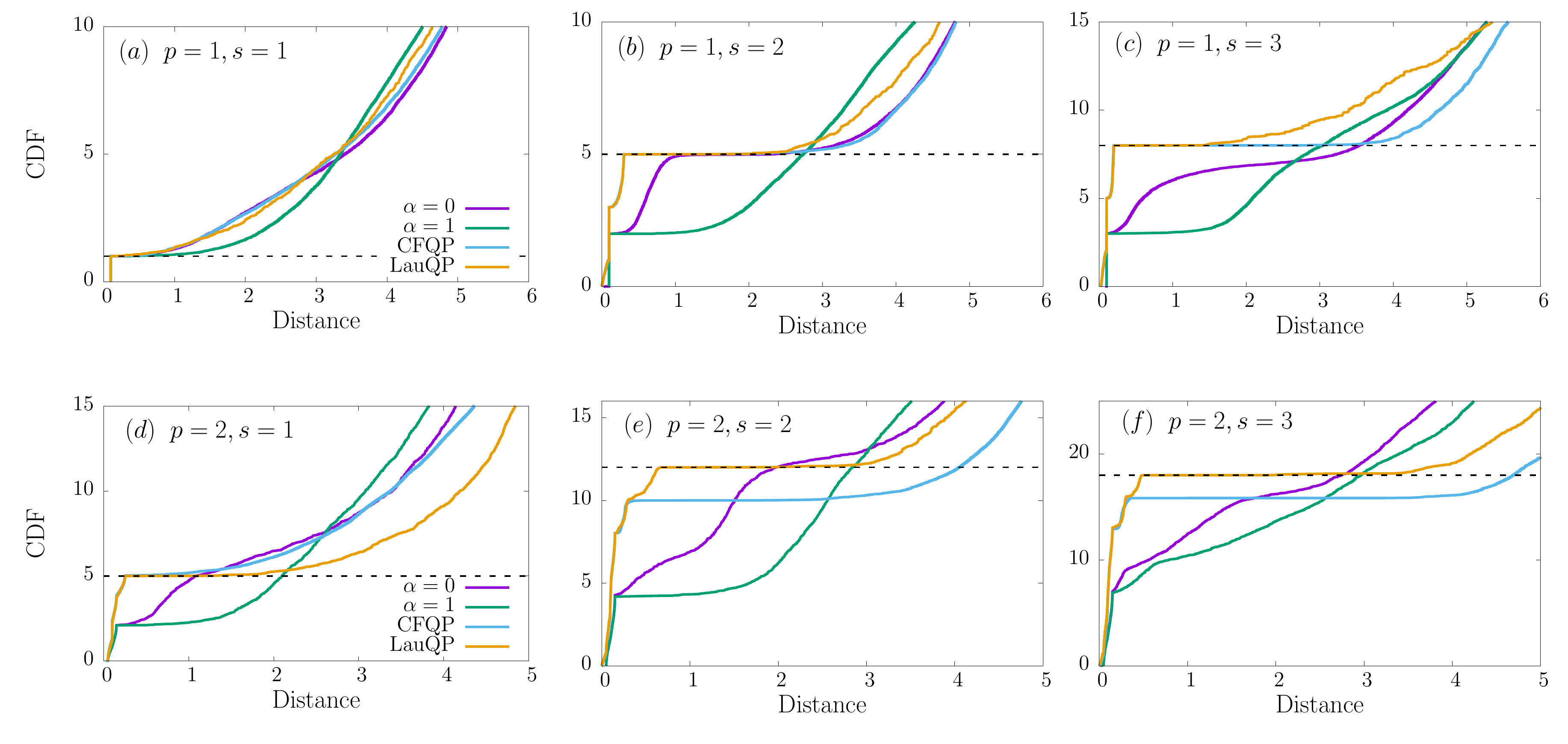} 
\caption{ CDFs of zeros in the QP states at $\nu =1/3$ for $N=9$ particles at a total angular momentum $L_{\rm total}=2N(N-1)/2+(N-1)(N-2)/2-1=99$ and maximum single-particle angular momentum $m_{\rm max}=3(N-1)-1=23$. CDFs for Laughlin QP and CF QP trial wavefunctions, and the ground states of pseudopotential states. Dashed lines are expected number of zeros from the root partition $\lambda_{\rm QP}^{1/3}$  using the formula $\gamma_{p+s}-\gamma_s-\gamma_p$.  }\label{p-1}
\end{figure*}

\begin{figure}
\includegraphics[width=0.45\textwidth]{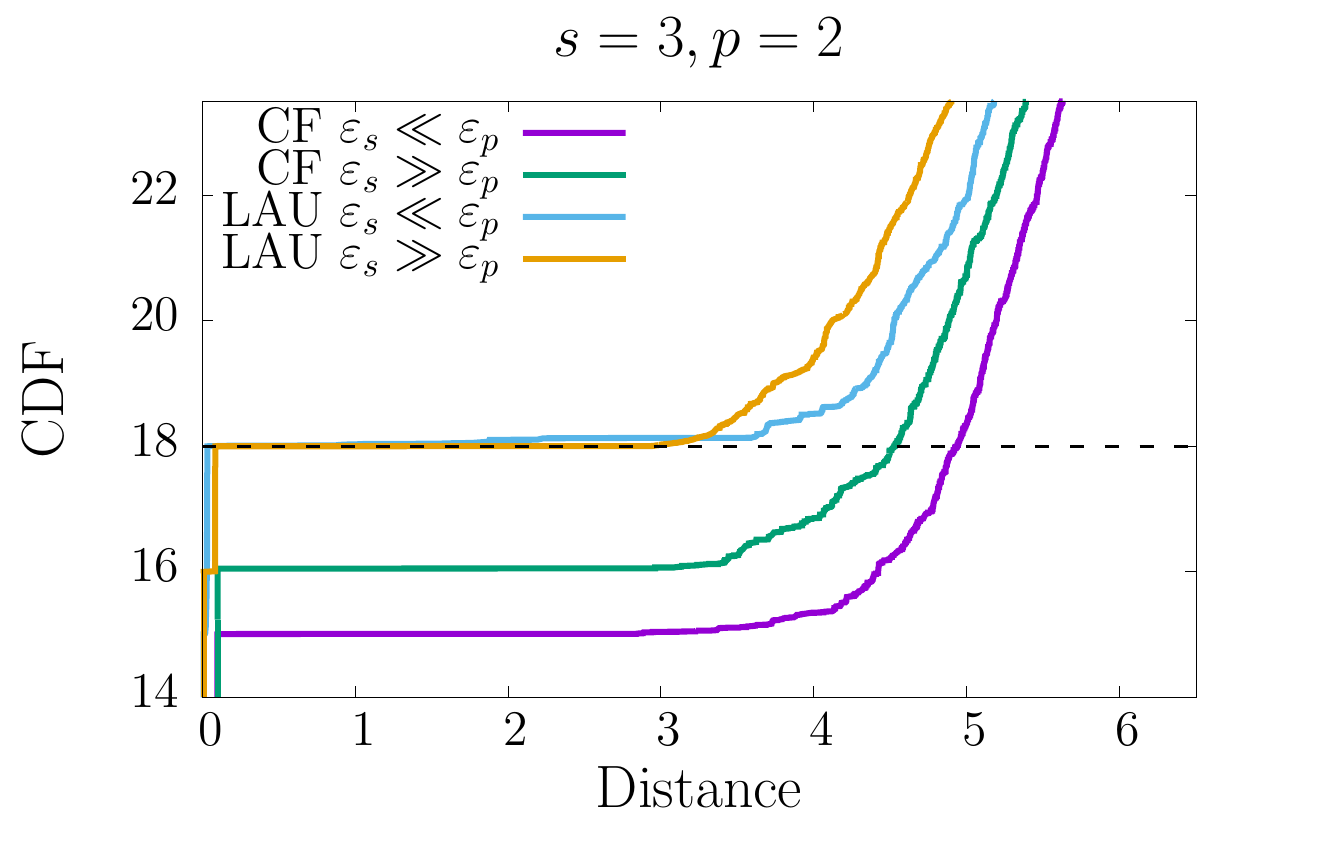} 
\caption{ CDFs of zeros in CF QP and  Lauglin QP trial wavefunctions at $\nu = 1/3$ for $N=9$ particles. Cluster sizes are $s=3, p=2$.  For each wave function two CDFs are presented, one for $\varepsilon_s \gg \varepsilon_p$ and another for $\varepsilon_s \ll \varepsilon_p$. The values of plateaus for CF trial wavefunction match the calculations in the Sec.~\ref{CFzeross}.
}\label{nmfig}
\end{figure}

The scaling proportional to $\sqrt{\alpha}$ can be understood as follows. Consider positions of zeros near a particle $z_1$ as seen by another particle say $z_N = x$.  
At $\alpha=0$, there are three overlaping zeros at $z_1$, which implies, that 
$\Psi(z_1,z_2,...,z_{N-1},x) = (z_1-x)^3\times Q (z_1,...,x)$, where $Q$ carries the information about the zeros far away from $z_1$. 
When the $V_1$ Hamiltonian is perturbed towards $V_C$, where $\alpha \ll 1$  the ground state of the perturbed interaction has a form
\begin{equation*}
\begin{split}
\Psi_{\alpha}(z_1,...z_N) = (z_1-x) \left[ (z_1-x)^2 +\alpha R(z_1-x) \right ] \times\\
\times Q'(\alpha,z_1,z_2,...x).
\end{split}
\end{equation*}
where $R(z)$ is a polynomial which depends on the locations of all other particles. The above term ensures one Pauli zero at $z_1$. For general locations of the particles, $R$ should have a constant term $C$. Considering this, the two non-Pauli zeros evolve as 
\begin{equation*}
    x = z_1 \pm \sqrt{-C\alpha}.
\end{equation*}
which explains both the power law scaling and zeros drifting in the diametrically opposite directions in a symmetric manner at small $\alpha$.

\subsection{Cumulative distribution functions}\label{ref4furt}


\begin{figure*}[ht]
\includegraphics[width=0.95\textwidth]{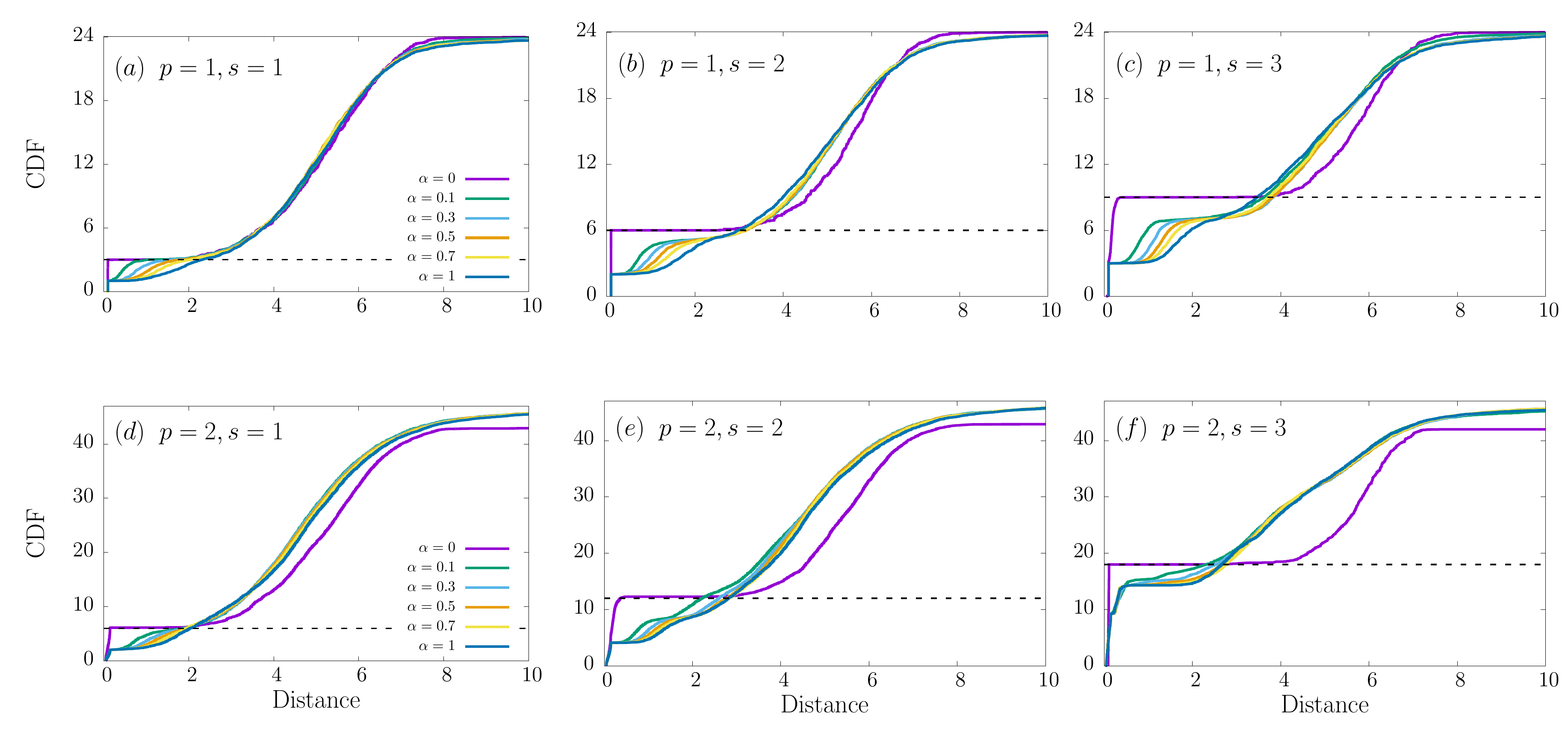} 
\caption{  CDFs of zeros in the ground states of pseudopotential states at $\nu = 1/3$. System size is $N=9$ particles at a total angular momentum $L_{\rm total}=3N(N-1)/2=108$ and maximum single-particle angular momentum $m_{\rm max}=3(N-1)=24$. Dashed lines are expected number of zeros from the root partition $(0,3,6,9...)$  using the formula $\gamma_{p+s}-\gamma_s-\gamma_p$. }\label{fig:gsL}
\end{figure*}

As discussed in the previous section, the three zeros associated with each particle in the ground state of $V^\alpha$ do not sit on the particles. The zeros between clusters do sit on the center of the cluster even in the limit of $\varepsilon_s$ and $\varepsilon_p$ approaching $0$. However, the number of zeros may still be in agreement with that of the trial wavefunctions in an approximate sense, as these zeros may stay in the close proximity of the particles or the clusters they are associated with. 
In order to test this, we consider the cumulative distribution of zeros in the vicinity of the clusters.

We sample configurations that have a cluster of $s$-particles in the vicinity of the origin as described in Sec.~\ref{nummet}. We then find the zeros seen by a $p$-cluster by solving for the roots of the reduced polynomial $P(x)$. The cumulative radial distribution of these zeros shows the number of zeros within a circle of radius $r$ around the origin. The distribution is averaged over several configurations. Such a distribution provides a way to see how the number of zeros attached to clusters gradually changes as the interaction deviates from the $V_1$ interaction.

\subsubsection{Cumulative distributions in the QP trial wavefunctions}

We first provide the cumulative distribution of the zeros in the QP wavefunctions with the QP located at the origin. These are shown in Fig.~\ref{p-1}. The top row shows the distribution of zeros for cases where there is $p=1$ particle in the $p$-cluster and $s=1,2,3$ particles in the $s$-clusters. We see that $s=1,2,3$ particle clusters are associated with $1$, $5$, and $8$ zeros respectively in both the CF and the Laughlin QP wavefunctions. These numbers match the root partition sequence (Eq.~\ref{parti1}) for these states.
The cumulative distributions show a plateau until a finite radius indicating an absence of any other zeros within a region around the origin. 
Note that the small deviations seen at very short radii are related to the finiteness of $\varepsilon_s$ used in the numerical calculations. These deviations vanish if a smaller $\varepsilon_s$ is chosen for the computation. 

The bottom row of Fig.~\ref{p-1} shows the similar results when the $p$-cluster has more than $1$ particles. As discussed in Sec.~\ref{CFzeross}, the number of zeros seen by a $p$-cluster ($p\geq 2$) on an $s$-cluster ($s\geq 2$) can sensitively depend on the manner in which $\varepsilon_s$ and $\varepsilon_p$ are sent to zero. In these plots we consider the path $\varepsilon_s=\varepsilon_p\to 0$. 

A simple counting based on the root partition suggests $\gamma_{p+s}-\gamma_p-\gamma_s$. From the cumulative distributions shown in Fig.~\ref{nmfig}, we empirically find that the number of the zeros in the Laughlin QP wavefunction matches this number. On the other hand, corresponding counting in CF QP wavefunction matches the counting derived in Sec.~\ref{CFzeross} and not equation given earlier in this paragraph. 

We now compare these results for the trial wavefunctions with the wavefunctions obtained from the exact diagonalization. We consider the lowest energy state obtained by diagonalization of $V^\alpha$ at a total angular momentum $L_{\rm total}=2N(N-1)/2+(N-1)(N-2)/2-1=99$ and maximum single-particle angular momentum $m_{\rm max}=3(N-1)-1=23$. These quantum numbers match that of the QP trial wavefunctions.
Note that even in the case of the exact parent Hamiltonian $V^{\alpha=0}$, does not generate the trial wavefunctions. 

The cumulative distributions for the eigenstate of  the $V_1$ interaction ($\alpha=0$) and the Coulomb interaction ($\alpha=1$) are shown in majenta and green colors in the Fig.~\ref{p-1}.
In the case of simple clusters $s=1,2$ and $p=1$, we find a match between trial wavefunctions and the $\alpha=0$ eigenstate. 
For larger clusters, $s\geq 2, p\geq 2$, the non-Pauli zeros are pushed away, and we do not find a clear plateau in the cumulative distributions to suggest any agreement with trial wavefunctions. At $\alpha=1$, only the Pauli zeros are associated with the clusters and we do see any plateau that resembles the ones in the trial wavefunctions.

In Fig.~\ref{nmfig}, we consider the cumulative distribution of the zeros seen by a $p=2$ cluster around an $s=3$ particle cluster. Two cases are analysed here, $\varepsilon_s\ll \varepsilon_p$ and $\varepsilon_p\ll \varepsilon_s$. In the Laughlin QP state, we find that the two cases produce an identical number of zeros whereas, in the CF QP wavefunction, the number of zeros depends on the relative size of the clusters. The numbers are consistent with the results obtained in Sec.~\ref{CFzeross} where we showed that the number of zeros depends on the order in which the two limits $\varepsilon_s\to0$ and $\varepsilon_p \to 0$ are taken.





\subsubsection{Cumulative distributions in the ground state of $V^\alpha$}

Having discussed the cumulative distribution of the zeros in the QP trial wavefunctions, we now explore how this distribution varies as the interaction $V^\alpha$ is tuned. These results are shown in Fig.~\ref{fig:gsL}. At all values of $V^\alpha$, the ground state is obtained for the total angular momentum sector $L_{\rm total}=3N(N-1)/2=108$ and in the Hilbert space where the maximum single-particle angular momentum $m_{\rm max}=3(N-1)=24$. These numbers match the ones for the Laughlin state.

For $\alpha =0$, the distribution of zeros seen by a $p=1$ cluster matches the expected number of zeros near the origin ($3,6,9$ zeros for  $s=1,2,3$ particle clusters respectively). For $\alpha>0$, there is a range of distances around the $s=1$ particle cluster where a plateau of $3$ is present. This is consistent with the observation made in Fig.~\ref{fig:gsL} that three zeros are still associated with each particle though two of them are located at a small finite distance from the particle.

However, for $s$-clusters of size $s>1$, we find less than the expected number of zeros. From the root partition, we expect that a $p=1$ cluster should see $6$ and $9$ zeros in the vicinity of  $s=2$ and $s=3$ particle cluster respectively. Instead, we find $5$ and $7$ zeros in these cases. 
Even in the case of $p>1$, we find less than the expected number of zeros surrounding the $s$-cluster.

Note that the cumulative distribution saturates to a constant at large radii. This saturation value indicates the total number of zeros seen by the $p$-clusters. This quantity should match the degree of the reduced polynomial $P(x)$. For the case of $p=1$ cluster, this degree is the largest allowed single-particle angular momentum $3(N-1)$. This is independent of $\alpha$ and consequentially all curves in the top row of Fig.~\ref{p-1} converge to the same value.

In the case of the $p>1$, the degree of the reduced polynomial is dependent on the dominance of partitions occurring in the expansion of the wavefunction. For $\alpha=0$, which produces the Laughlin state as the ground state, the fact that the root partition is $(0,3,6\dots)$ ensures that the degree of the reduced polynomial is $3\sum_{i=1}^p (N-i)$. At finite $\alpha$, in general, all basis states have a finite amplitude, including the basis states labeled by partitions that are not dominated by the Laughlin root partition. The degree of the reduced polynomial is then given by the maximum possible total angular momentum of $p$ particles which is $\sum_{i=1}^{p} 3(N-1)-i+1$.

\section{Conclusions}
Analysis of quantum Hall wavefunctions in terms of the root partitions (or closely related notions) which are associated with the zeros seen by a particle on a cluster of n other particles has been attempted in the past  \cite{Bernevig08,Bandyopadhyay18, Sreejith18,Regnault09,Xiao008,PhysRevB.79.195132,PhysRevB.100.241302,Bernevig08,Bernevig08a,Bernevig09,BARATTA2011362,Lee15,Kusmierz2016QuantumHS}. Here, we considered a straightforward generalization of this notion, wherein we ask about the number of zeros $\gamma_{p,s}$ seen by a $p$-particle cluster on an $s$ particle cluster. The root partition itself can be associated with the sequence of numbers $\gamma_{p=1,s}$.

In clustered states with simple root partitions like the Laughlin state and the Pfaffian \cite{Xiao008}, the numbers $\gamma_{p,s}$ are given through the relation $\gamma_{p,s} = \sum_{i=1}^{p+s} \lambda_i-\sum_{i=1}^{p} \lambda_i- \sum_{i=1}^{s} \lambda_i$ where $\lambda$ is the corresponding root partition ($(0,3,6...)$ and $(0,0,2,2,4,4,...)$ for the Laughlin and Pfaffian states respectively).

It is natural to ask if the $\gamma_{p,s}$ is uniquely determined in a similar manner by general composite fermion trial wavefunctions at filling $n/(2pn+1)$. Addressing this analytically or computationally is a challenging task. We instead considered the simplest states that go beyond the Laughlin state, namely the single QP trial wavefunctions.

Two different trial wavefunctions have been considered for such a system, namely the Laughlin QP wavefunction and the CF QP wavefunction. They both share the same root partition. However, we find that in the CF QP wavefunction, $\gamma_{p,s}$ is not uniquely determined by the root partition on account of the fact that the number of zeros attached to a cluster depends on the radii of the $p$ and $s$ clusters. Numerical studies on small $p,s$ indicated that $\gamma_{p,s}$ for the Laughlin QP wavefunction, on the other hand, is uniquely determined by the root partition by the above mentioned relation.

We further considered the question of how well the root partition and the more general zero counting properties of the trial wavefunctions are inherited by the eigenstates of Hamiltonians in the same universality class. We considered the ground states and the single QP states of the $V_1$ interaction (parent Hamiltonian of the Laughlin state) with an added Coulomb perturbation. Even for very small perturbations away from the parent Hamiltonian, the zeros attached to clusters decouple, and the counting of zeros on the clusters deviate from the corresponding numbers in the trial functions. 
The counting of zeros implies absence of certain angular momentum channels in the state. These angular momentum constraints appear to be obeyed even in the Coulomb ground state, not just in the trial states (See Table V of Ref.~\onlinecite{Sreejith18}).
These results indicate that angular momentum constraints maybe a more robust at characterising FQH phases than the root partition.

\bibliography{main}
\appendix

\section{Unprojected CF QP states and ``infinitelly'' many zeros} \label{unprr}
Number of zeros of the unprojected CF QP trial wavefunction $\Psi_{\rm CB}^{\rm QP,unp}$ at filling $\nu = 1/r$ with $p=1$ particle in the $p$-cluster can vary from $r(N-1)$  to  ``infinitely many''. This is apparent when the wave function is written using the determinant 
\begin{equation}
\Psi_{\rm CB}^{\rm QP,unp}={ \left| \begin{array}{ccccc}
\bar{z}_1 & \bar{z}_2 & . &.&.\\
1 & 1 & . &.&. \\
z_{1}&z_{2}&. &.&.\\
.&.&.&.&.\\
.&.&.&.&.  \\
z_1^{N-2} & z_2^{N-2} & . &.&.
\end{array} \right | } \times  \prod_{i<j} (z_i -z_j)^{r-1}. 
\end{equation}
For $p=1,s=N-1,f=0$ the Jastrow factor always assures that the wave function vanishes as $(r-1)^{\rm{ th}}$ power when two particles meet. Moreover, the determinant, assures an additional zero on each particle when the position of $p$-cluster coincides with the particle (two matrix columns are the same). 

We note that in certain symmetrical configurations of particles there are infinitely many points at which the wavefunction vanishes. Let us fix the positions of particles in the $s$-cluster to lay in a straight line {\emph{i.e.}} $\varepsilon_s^{(i)}=r_i\times c$ for some $r_i \in \mathbb{R}$ and complex constant $c \in \mathbb{C}$. Then for any position of probing cluster on this line $x = r_p\times c $ ($r_p \in \mathbb{R}$) the determinant term vanishes, which is a consequence of the first and third row of the matrix are proportional with the proportionality ratio ${c}/{\bar{c}}$.

\section{Particle-Zeros correlations}\label{app:standcf}

\begin{figure}[ht]

\includegraphics[width=0.45\textwidth]{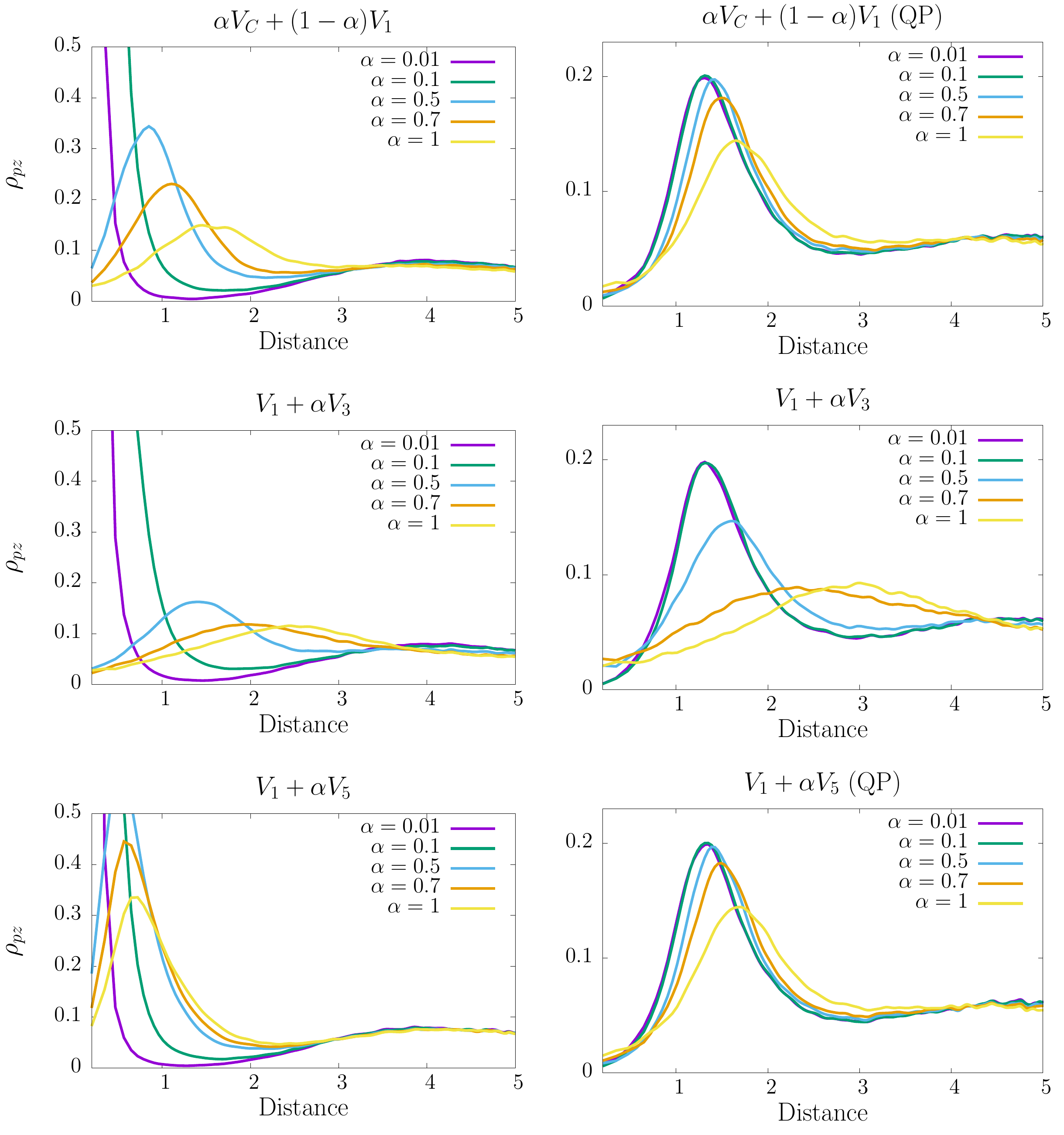} 
\caption{The particle-zero correlation function $\rho_{pz}(r)$ for the exact diagonalisation ground states at the angular momenta corresponding to the Laughlin ground and QP states at $\nu=1/3$. Three interactions $\alpha V_C + (1-\alpha)V_1$, $V_1+\alpha V_3$ and  $V_1+\alpha V_5$ are presented. Pauli zeros are not included.}\label{notrack2}
\end{figure}

The particle-zero correlation function $\rho_{pz}(r)$ is the conditional probability density of finding a zero-particle pair at the distance $r$. In Fig.~\ref{notrack2}  we provide them for the exact diagonalisation ground states at the angular momenta corresponding to the Laughlin ground and QP states at $\nu=1/3$. Three perturbations of $V_1$ are considered, namely $\alpha V_C + (1-\alpha)V_1$, $V_1+\alpha V_3$ and  $V_1+\alpha V_5$.

\end{document}